\documentclass[11pt,letterpaper]{article}

\pdfoutput=1 
\usepackage{color}
\usepackage{jcappub}
\usepackage{amsmath}
\usepackage{amsfonts}
\usepackage{amssymb}
\usepackage{graphicx}
\usepackage{epsfig}
\usepackage{natbib}

\usepackage{subcaption}
\usepackage[thinlines]{easytable}
\usepackage{array}
\usepackage[font=small]{caption}
\def\ba{\begin{eqnarray}}
\def\ea{\end{eqnarray}}
\def\bea{\begin{eqnarray}}
\def\eea{\end{eqnarray}}
\def\be{\begin{equation}}
\def\ee{\end{equation}}

\def\({\left(}
\def\){\right)}
\def\[{\left[}
\def\]{\right]}

\newcolumntype{M}[1]{>{\centering\arraybackslash}m{#1}}

\title{Modifications to Cosmological Power Spectra from Scalar-Tensor Entanglement and their Observational Consequences}
\author[a]{Nadia Bolis,}
\author[a]{Andreas Albrecht,}
\affiliation[a]{Department of Physics, University of California at Davis, One Shields Ave, Davis, CA 95616, USA}
\author[b,c]{R.~Holman}
\affiliation[b]{Physics Department, Carnegie Mellon University, Pittsburgh, PA 15213, USA}
\affiliation[c]{College of Computational Sciences, Minerva University, 1145 Market Street, San Francisco, CA 94103, USA}
\emailAdd{nbolis@ucdavis.edu}
\emailAdd{ajalbrecht@ucdavis.edu}
\emailAdd{rh4a@andrew.cmu.edu}

\abstract{We consider the effects of entanglement in the initial quantum state of scalar and tensor fluctuations during inflation. We allow the gauge-invariant scalar and tensor fluctuations to be entangled in the initial state and compute modifications to the various cosmological power spectra. We compute the angular power spectra ($C_l$'s) for some specific cases of our entangled state and discuss what signals one might expect to find in CMB data. This entanglement also can break rotational invariance, allowing for the possibility that some of the large scale anomalies in the CMB power spectrum might be explained by this mechanism.}

\begin{document}

\maketitle

\section{Introduction\label{sec:intro}}
We now have a great deal of information about the power spectrum \cite{Aghanim:2015xee} and to a lesser extent, about the bi-spectrum of CMB anisotropies \cite{Ade:2015ava}. These are consistent with what inflation would predict {\em if} the quantum state of inflaton fluctuations was chosen to be the Bunch-Davies (BD) \cite{Bunch:1978yq} state. There are two ways to interpret this. One is that, to the extent that we do believe that quantum fluctuations of the inflaton are indeed responsible for the temperature anisotropies in the CMB, we have been given the directive that nature chooses to use the nearest thing to a vacuum state that a nearly de Sitter, inflationary universe allows. The other is to adopt a more skeptical point of view and ask to what extent are other states truly ruled out by the data. This latter viewpoint has been used by a number of authors who considered corrections to the power spectrum \cite{Martin:2000xs,Danielsson:2002kx,Kaloper:2002uj,Collins:2005nu,Collins:2006bg,Carney:2011hz}, as well as to the bi-spectrum \cite{Chen:2006nt,Holman:2007na,Meerburg:2009ys,Agullo:2010ws,Ashoorioon:2010xg,Ganc:2011dy} from the use of excited states based on the BD state. More general states, such as mixed ones \cite{Agarwal:2012mq}, non-Bunch Davies vacuum state \cite{Dey:2012qp,  Kundu:2011sg, Kundu:2013gha, deAlwis:2015ioa} and correlated causally disconnected regions \cite{Kanno:2015lja} have also been considered. An interesting and widely discussed specific case where the initial state of inflation can be non-BD is  when inflation for our observed ``pocket universe" starts with a tunneling event (as discussed for example in \cite{Albrecht:2011yg}). This could lead to interesting observable phenomena if the inflation within the pocket universe is sufficiently short.
More recently, a new class of states has been examined, one in which the inflaton is {\em entangled} with another scalar field \cite{Albrecht:2014aga}. This entanglement modifies the power spectrum by introducing oscillatory features that depend on the mass and coupling to gravity (minimal or conformal) of the other fields. In this work, we continue our examination of entangled states by considering states in which the gauge invariant scalar and {\em tensor} perturbations are entangled with each other. 

Generally our motivation is driven by the possibility that the EFT that describes inflation may be emergent from some more fundamental theory right at the start of inflation (as considered for example in  \cite{Albrecht:2014eaa}).  We do not have a concrete description of this process, so we resort to a phenomenological framework that simply assumes the EFT emerges with a slightly more general form for the wavefunction than Bunch-Davies.  Another point of view one might take is that we are embracing hints from the data that there may be a small breaking of rotational invariance in the state of the universe, and are considering a simple extension of Bunch-Davies that allows for this sort of breaking. The state we study here is an entangled Gaussian which is the next to simplest state to a Gaussian considering possible evidence for rotational invariance breaking in the data. The non-trivial transformations of the tensor perturbations under the rotation group now allow for the breaking of rotational invariance; such breaking is constrained by current data, but might still be large enough to explain some of the large scale anomalies \cite{Ade:2015hxq} found in the CMB temperature anisotropy maps. 

In the next section we set up the entangled initial state and evolve it using the Schr\"{o}dinger picture formalism. We then compute the various power spectra produced by these states. Since the scalar and tensor perturbations transform differently under rotations, some of the standard relations, such as the fact that $\langle a_{l m} a^*_{l^{\prime} m^{\prime}}\rangle \propto \delta_{l l^{\prime}}$ no longer hold. We compute the angular power spectra $C_l$'s for different magnitudes of entanglement and discuss how our results should be compared to existing data. We also discuss to what extent these states might explain any or all of the large scale anomalies mentioned above.

In a companion paper \cite{Collins:2016ahj} a different way to construct such entangled states is described which makes use of the Lagrangian oriented formalism developed in refs. \cite{Agarwal:2012mq} and \cite{Collins:2013kqa}.

\section{\label{sec:entangstate} Entangling Scalar and Tensor Perturbation Modes}

\subsection{\label{subsec:schrodinger} The Schr\"{o}dinger Picture Approach}

In order to describe the entanglement between the scalar perturbation $\zeta$ and the tensor perturbations $\gamma_{i j}$, we use  Schr\"{o}dinger picture field theory \cite{Boyanovsky:1993xf,Anderson:2005hi,Freese:1984dv} in this subsection (though for another viewpoint on the states constructed in ref.\cite{Albrecht:2014aga} see ref.\cite{Kanno:2014ifa}). This entails constructing the Hamiltonian for the $\zeta$-$\gamma_{i j}$ system as well as giving the wave-functional $\Psi[\zeta, \gamma_{i j}]$ which will solve the Schr\"{o}dinger equation coming from the Hamiltonian. 


 We consider the case where the quadratic parts of the action for $\zeta$ and $h_{i j}$ dominate and thus, at this level, $\zeta$ and $\gamma_{i j}$ are decoupled in the action:

\begin{equation}
\label{eq:action}
S_{\rm quad} = \int d^4x a^3(t) \left[\frac{\epsilon M^2_{pl}}{2} (\partial_{\mu} \zeta \partial^{\mu} \zeta) + \frac{M_{pl}^2}{8} \partial_{\mu} \gamma_{ij}\partial^{\mu}\gamma^{ij}\right],
\end{equation}
where $\epsilon$ is the slow roll parameter, $M_{pl}$ the Planck mass, and $a(t)$ the scale factor. If we go to conformal time $\tau$ and decompose the tensor perturbation into the polarization basis, then the action takes the form
\begingroup\makeatletter\def\f@size{10}\check@mathfonts\begin{equation}
\label{eq:conftimeaction}
S_{\rm quad} = \int d^4x \frac{a(\tau)^2 M^2_{pl}}{2} \left[\epsilon\left(\zeta^{\prime}(\tau, \vec{x})^2 - (\nabla \zeta(\tau, \vec{x}))^2\right) + \frac{1}{2} \sum_{\sigma=+,\times}\left(h^{(\sigma) \prime}(\tau, \vec{x})^2-\left(\nabla h^{(\sigma)}(\tau, \vec{x})\right)^2 \right)\right],
\end{equation}
\endgroup
where primes denote conformal time derivatives and we have defined $h^{(\sigma)}$ via:

\begin{equation}
\label{eq:polbasis}
\gamma_{i j} = \sum_{\sigma=+,\times} e_{ i j}^{(\sigma)} h^{(\sigma)},\quad {\rm with}\quad e_{i j}^{(\sigma)} e^{(\sigma^{\prime})\ i j}=4 \delta^{\sigma \sigma^{\prime}}.
\end{equation}
We then find the Hamiltonian for the system in the usual way by first computing the conjugate momenta for both scalar and tensor modes:
\begin{equation}
\label{eq:canmom}
\Pi = \frac{\delta \mathcal{L}}{\delta \zeta^{\prime}} =  \epsilon a^2 M^2_{pl}  \zeta^{\prime}, \quad \Pi^{(\sigma)} = \frac{\delta \mathcal{L}}{\delta h^{(\sigma) \prime}} = a^2 M^2_{pl} h^{(\sigma) \prime}.
\end{equation}
Using eq.(\ref{eq:canmom}), the Hamiltonian then takes the form,
\begin{equation}
\label{eq:hamiltonian}
H = \int d^3x \left[ \frac{\Pi^2}{2 a^2 \epsilon M_{pl}^2} + \frac{a^2 \epsilon M_{pl}^2}{2} (\nabla \zeta)^2 +\sum_{\sigma=+,\times}\left(\frac{\Pi^{(\sigma) 2}}{2 a^2 M_{pl}^2} +  \frac{a^2 M_{pl}^2}{2} (\nabla h^{(\sigma)})^2\right)\right].
\end{equation} 
For later convenience, we define $\alpha^2 =  a^2 \epsilon M_{pl}^2$ and $\beta^2 = a^2 M_{pl}^2$. We will also use the spatial flatness of the FRW spacetime to decompose both $\zeta$ and $h^{(\sigma)}$ in terms of their respective (box normalized) momentum modes; here $V$ is the comoving spatial volume of the box used in the normalization:
\begin{eqnarray}
\label{eq:mommodes}
\zeta(\vec{x}) = \sum_{\vec{k}} \frac{\zeta_{\vec{k}}}{\sqrt{V}} e^{- i \vec{k} \cdot \vec{x}} &,& \quad \Pi(\vec{x}) = \sum_{\vec{k}} \frac{\Pi_{\vec{k}}}{\sqrt{V}} e^{- i \vec{k} \cdot \vec{x}}\nonumber, \\
h^{(\sigma)}(\vec{x}) = \sum_{\vec{k}} \frac{h^{(\sigma)}_{\vec{k}}}{\sqrt{V}} e^{- i \vec{k} \cdot \vec{x}}&,& \quad \Pi^{(\sigma)}(\vec{x}) = \sum_{\vec{k}} \frac{\Pi^{(\sigma)}_{\vec{k}}}{\sqrt{V}} e^{- i \vec{k}\cdot \vec{x}}.
\end{eqnarray}
The fact that the Hamiltonian is quadratic in the fields allows the different momentum modes to decouple from each other so that the Hamiltonian decomposes into a sum of separate Hamiltonians for each mode:
\begin{equation}
H = \sum_{\vec{k}} (H_{\zeta \vec{k}} + \sum_{\sigma=+,\times}H^{(\sigma)}_{\vec{k}}),
\end{equation}
with the Hamiltonians of $\zeta$ and $\gamma_{ij}$ respectively, 
\begin{eqnarray}
\label{eq:modehams}
H_{\zeta \vec{k}} &=& \frac{\Pi_{\vec{k}} \Pi_{-\vec{k}}}{2 \alpha^2} + \frac{k^2 \alpha^2}{2} \zeta_{\vec{k}} \zeta_{-\vec{k}}, \\ 
H^{(\sigma)}_{\vec{k}} &=& \frac{\Pi^{(\sigma)}_{\vec{k}} \Pi^{(\sigma)}_{-\vec{k}}}{2 \beta^2} + \frac{k^2 \beta^2}{2} h^{(\sigma)}_{\vec{k}}h^{(\sigma)}_{-\vec{k}}.
\end{eqnarray}
In the Schr\"{o}dinger picture the state is represented with a wave-functional of the field modes $\Psi[\{\zeta_{\vec{k}}\}, \{h^{(\sigma)}_{\vec{k}}\}, \tau] $ which obeys the functional Schr\"{o}dinger equation:

\begin{equation}
\label{eq:functSeqn}
i\partial_{\tau} \Psi[\{\zeta_{\vec{k}}\},\{h^{(\sigma)}_{\vec{k}}\}, \tau] = H \Psi[\{\zeta_{\vec{k}}\},\{h^{(\sigma)}_{\vec{k}}\}, \tau],
\end{equation}
where the momenta become differential operators as usual:

\begin{equation}
\Pi_{\vec{k}}=-i\frac{\delta}{\delta \zeta_{-\vec{k}}},\quad \Pi^{(\sigma)}_{\vec{k}}=-i\frac{\delta}{\delta h^{(\sigma)}_{-\vec{k}}}.
\end{equation}
In the absence of any interactions in the Hamiltonian it is consistent to factorize the wave-functional into a product of wave-functions for each momentum mode: 
\begin{equation}
\Psi[\{\zeta_{\vec{k}}\}, \{h^{(\sigma)}_{\vec{k}}\}, \tau] = \prod_{\vec{k}} \psi_{\vec{k}}[\zeta_{\vec{k}}, h^{(\sigma)}_{\vec{k}} , \tau].
\end{equation}
We take the wave-functions for each mode to be Gaussians
\begin{eqnarray}
\label{eq:wavefunctions}
&& \psi_{\vec{k}}[\zeta_{\vec{k}}, h^{(t)}_{\vec{k}}, \tau] =\nonumber \\
&& \sqrt{ N_{k}(\tau)} \exp{\left[-\frac{1}{2} \left(A_{k}(\tau) \zeta_{\vec{k}}\zeta_{-\vec{k}} + B^{(\sigma \sigma^{\prime})}_k(\tau) h^{(\sigma)}_{\vec{k}} h^{(\sigma^{\prime})}_{-\vec{k}} + C^{(\sigma)}_k(\tau) \left(\zeta_{\vec{k}} h^{(\sigma)}_{-\vec{k}} + \zeta_{-\vec{k}}h^{(\sigma)}_{\vec{k}}\right)\right)\right]},\nonumber\\
\end{eqnarray}
such that kernel $C^{(\sigma)}_k(\tau) \neq 0$ sets the entanglement between tensor and scalar modes. Note that we are summing over $\sigma, \sigma^{\prime}=+,\times$ in the above wave function and that we have allowed for non-diagonal couplings in the kernel $B^{(\sigma \sigma^{\prime})}_k(\tau)$ between the $+$ and $\times$ polarization modes. The full state for momentum $\vec{k}$ is the product $\psi_{\vec{k}}\ \psi_{-\vec{k}}$:
\begin{eqnarray}
\label{eq:fullstate}
&& \psi_{\vec{k}}\ \psi_{-\vec{k}}=\nonumber\\
&& N_{k}(\tau) \exp{\left[- \left(A_{k}(\tau) \zeta_{\vec{k}}\zeta_{-\vec{k}} + B^{(\sigma^{\prime} \sigma)}_{k S}(\tau) h^{(\sigma)}_{\vec{k}} h^{(\sigma^{\prime})}_{-\vec{k}} + C^{(\sigma)}_k(\tau) \left(\zeta_{\vec{k}} h^{(\sigma)}_{-\vec{k}} + \zeta_{-\vec{k}}h^{(\sigma)}_{\vec{k}}\right)\right)\right]},\nonumber\\
\end{eqnarray}
where $B^{(\sigma^{\prime} \sigma)}_{k S}(\tau)$ is the symmetric part of $B^{(\sigma \sigma^{\prime})}_k(\tau)$, where we have used the fact that since both $\zeta$ and $h^{(+,\times)}$ are real fields, we know that $\zeta_{-\vec{k}}= \zeta^*_{\vec{k}}$ and likewise for $h^{(\sigma)}_{\vec{k}}$. For simplicity we'll define the matrix $\left(\mathbb{B}_k\right)^{(\sigma \sigma^{\prime})}\equiv B^{(\sigma \sigma^{\prime})}_{k S}(\tau)$.
The functional Schr\"{o}dinger equation, eq.(\ref{eq:functSeqn}), factorizes into an infinite number of ordinary Schr\"{o}dinger
equations, one for each mode:
\begin{equation}
i \partial_{\tau} \psi_{\vec{k}}[\zeta_{\vec{k}},h^{(\sigma)}_{\vec{k}}, \tau]  = \left( H_{ \zeta \vec{k}} + \sum_{t=+,\times}H^{(t)}_{\vec{k}}  \right) \psi_{\vec{k}}[\zeta_{\vec{k}}, ,h^{(\sigma)}_{\vec{k}}, \tau] .
\end{equation}
Inserting our Gaussian anzatz gives us equations of motion for the kernels $A_k(\tau)$, $B^{(s s^{\prime})}_{k S}(\tau)$, $ C^{(\sigma)}_k(\tau)$ and the normalization factor $N_k(\tau)$,
 \begin{eqnarray}
 \label{eq:stateeqns}
 i \frac{N^{\prime}_k}{ N_k} &=& \left(\frac{A_k}{\alpha^2} + \frac{{\rm Tr}(\mathbb{B}_{k})}{\beta^2} \right), \nonumber\\
 i A^{\prime}_k &=& \frac{A_k^2}{\alpha^2} + \frac{(\mathbf{C}_k^{T} \mathbf{C}_k)}{\beta^2} - \alpha^2 k^2, \nonumber \\
 i \mathbb{B}_{k}^{\prime}&=& \frac{\mathbb{B}_{k}^2}{\beta^2} + \frac{(\mathbf{C}_{k}\mathbf{C}_k^{T})}{\alpha^2} - \beta^2 k^2 \mathbb{I}, \nonumber \\
 i \mathbf{C}_k^{\prime}&=& \left(\frac{A_k}{\alpha^2}\ \mathbb{I} + \frac{\mathbb{B}_{k}}{\beta^2} \right)\mathbf{C}_k , 
 \end{eqnarray}
 where we have defined the column vector $\left(\mathbf{C}_k\right)^{(\sigma)}\equiv C^{(\sigma)}_k(\tau)$. Note that from the third line in eq.(\ref{eq:stateeqns}), we see that if $\mathbb{B}_k$ is diagonal, then one of $C^+$ or $C^{\times}$ has to vanish so that $\mathbf{C}_{k}\mathbf{C}_k^{T}$ is also diagonal. In particular, if $\mathbb{B}_k$ is proportional to the identity, then {\em both} $C^+$ and $C^{\times}$ must be zero, forcing the scalar and tensor modes to disentangle themselves.
 
In order to solve eqs.(\ref{eq:stateeqns}), we write $\mathbb{B}_{k}$ in terms of the identity and the Pauli matrices, where since $\mathbb{B}_{k}$ is symmetric, we can omit $\sigma_2$ in this decomposition:
\begin{equation}
\mathbb{B}_{k} = b_{0 k}\mathbb{I}+ \vec{b}_k\cdot \vec{\sigma},\quad b_{2 k}\equiv 0,
\end{equation}
to find 
\begin{eqnarray}
\label{eq:paulistateeqns}
i \frac{N^{\prime}_k}{ N_k} &=& \frac{1}{2} \left(\frac{A_k}{\alpha^2} + \frac{2 b_{0 k}}{\beta^2} \right),\nonumber \\
 i A^{\prime}_k &=& \frac{A_k^2}{\alpha^2} + \frac{(\mathbf{C}_k^{T} \mathbf{C}_k)}{\beta^2} - \alpha^2 k^2, \nonumber  \\
 i b_{0 k}^{\prime} &=& \frac{b_{0 k}^2 + \vec{b}_k^2}{\beta^2} + \frac{(\mathbf{C}_{k}^T \mathbf{C}_k)}{2 \alpha^2} - \beta^2 k^2,\nonumber   \\
 i\vec{b}_k^{\prime} &=& \frac{2 b_{0 k} \vec{b}_k}{\beta^2} +\frac{(\mathbf{C}_{k}^T\ \vec{\sigma}\ \mathbf{C}_k)}{2 \alpha^2}, \nonumber  \\
 i \mathbf{C}_k^{\prime}&=& \left(\left[\frac{A_k}{\alpha^2} + \frac{b_{0 k}}{\beta^2}\right]\mathbb{I} + \frac{\vec{b}_k\cdot \vec{\sigma}}{\beta^2} \right)\mathbf{C}_k , 
\end{eqnarray}
The equations for $A_k$, $b_{0 k}$ are of the Ricatti form, so we can convert them 
into linear, second order equations by making the substitutions, 
 \begin{equation} 
 \label{eq:Ricatticonversion}
 i A_k = \alpha^2\left(\frac{f^{\prime}_k}{f_k} - \frac{\alpha^{\prime}}{\alpha}\right) , \quad  i b_{0 k} = \beta^2\left(\frac{g^{\prime}_k}{g_k} - \frac{\beta^{\prime}}{\beta}\right) ,
 \end{equation}
 leading to the following equations of motion for the mode functions $f_k(\tau)$ and $g_k(\tau)$:
 \begin{eqnarray}
 \frac{f^{\prime\prime}_k}{f_k} + \left(k^2 - \frac{\alpha^{\prime\prime}}{\alpha} \right) &=& \frac{\mathbf{C}_k^{T} \mathbf{C}_k}{\alpha^2 \beta^2}, \\
  \frac{g^{\prime\prime}_k}{g_k} + \left(k^2 - \frac{\beta^{\prime\prime}}{\beta} \right)&=& \frac{\mathbf{C}_k^{T} \mathbf{C}_k}{2 \alpha^2 \beta^2}+\frac{\vec{b}_k^2}{\beta^4}.
\end{eqnarray}
Note that as expected, the above equations imply we can take $b_{2 k}\equiv 0$ consistently. Also note that the equations for $\vec{b}_k^{\prime}$ and $\mathbf{C}_k^{\prime}$ admit integrating factors. Using eq.(\ref{eq:Ricatticonversion}) we can rewrite these equations in terms of the variables:

\begin{equation}
\vec{\tilde{b}}_k \equiv \frac{g_k^2}{\beta^2} \vec{b}_k\quad \tilde{\mathbf{C}}_k \equiv \frac{f_k g_k}{\alpha \beta} \mathbf{C}_k;
\end{equation}

\begin{eqnarray}
\label{eq:convertedeqns}
&& f^{\prime\prime}_k + \left(k^2 - \frac{\alpha^{\prime\prime}}{\alpha} \right) f_k = \frac{\tilde{\mathbf{C}}_k^{T} \tilde{\mathbf{C}}_k}{f_k g_k^2},\nonumber \\
&& g^{\prime\prime}_k + \left(k^2 - \frac{\beta^{\prime\prime}}{\beta} \right) g_k = \frac{\tilde{\mathbf{C}}_k^{T} \tilde{\mathbf{C}}_k}{2 f_k^2 g_k}+\frac{\vec{\tilde{b}}_k^2}{g_k^3},\nonumber\\
&& i\vec{\tilde{b}}_k^{\prime} = \frac{\tilde{\mathbf{C}}_k^T\ \vec{\sigma}\ \tilde{\mathbf{C}}_k}{2 f_k^2},\nonumber\\
&& i\tilde{\mathbf{C}}^{\prime}_k = \frac{\vec{\tilde{b}}_k\cdot \vec{\sigma}\ \tilde{\mathbf{C}}_k}{g_k^2} .
\end{eqnarray}
The quantity $\beta^{\prime\prime}/\beta$ is just equal to $a^{\prime\prime}/a$ while $\alpha^{\prime\prime}/\alpha$ becomes $\left(\frac{a^{\prime}}{a}\right)^2 \left[2 - \epsilon + \frac{3}{2} \eta \right]$ to first order in the slow roll parameters $\epsilon$ and $\eta$. 
Finally our equations of motion become
 \begin{eqnarray}
 \label{eq:finaleqns}
&& f^{\prime\prime}_k (\tau)+ \left(k^2 - \frac{\nu_{\zeta}^2 - \frac{1}{4}}{\tau^2} \right)f_k(\tau) = \frac{\tilde{\mathbf{C}}_k^{T} \tilde{\mathbf{C}}_k}{f_k g^2_k},\nonumber \\
&& g^{\prime\prime}_k (\tau)+ \left(k^2 - \frac{\nu_{\gamma}^2 - \frac{1}{4}}{\tau^2} \right)g_k(\tau) =\frac{\tilde{\mathbf{C}}_k^{T} \tilde{\mathbf{C}}_k}{f_k^2 g_k}+\frac{\vec{\tilde{b}}_k^2}{g_k^3},\nonumber\\
&& i\vec{\tilde{b}}_k^{\prime} = \frac{\tilde{\mathbf{C}}_k^T\ \vec{\sigma}\ \tilde{\mathbf{C}}_k}{2 f_k^2},\nonumber\\
&& i\tilde{\mathbf{C}}^{\prime}_k = \frac{\vec{\tilde{b}}_k\cdot \vec{\sigma}\ \tilde{\mathbf{C}}_k}{g_k^2} ,
\end{eqnarray}
with $\nu_{\gamma} = 3/2$ and $\nu_{\zeta} = \sqrt{\frac{3}{2} + \epsilon + \frac{1}{2} \eta} = \sqrt{\frac{3}{2}(1- n_s) + \frac{9}{4}}$ in terms of the spectral index $n_s$. 

As discussed above, the last two of eqs.(\ref{eq:finaleqns}) show that it is inconsistent to take both $\vec{\tilde{b}}_k = \vec{0}$ and $\mathbf{C}_k\neq \mathbf{0}$. The minimal consistent choices are either $\tilde{b}_{1 k}\neq 0,\ \tilde{C}^+=\pm \tilde{C}^{\times}$ (later referred to as \emph{Case 1}) or $\tilde{b}_{3 k}\neq 0$ with one of  $\tilde{C}^+,\ \tilde{C}^{\times}$ vanishing (\emph{Case 2}).

 \subsection{\label{sunsec:infstate} Normalizations and Two Point Functions}
 
Let's first consider under what circumstances is our state normalizable. Since the full state factorizes in the momentum label, we demand that the wave function for each momentum state be normalizable. Thus the condition for the wave functions to be normalizable is that
 
\begin{equation}
\label{eq:normalization}
\int {\cal D}^2 \zeta_{\vec{k}}\ \prod_{\sigma=+,\times} {\cal D}^2 h^{(\sigma)}_{\vec{k}}\ \left| \psi_{\vec{k}}\right|^2 \left| \psi_{-\vec{k}}\right|^2< \infty,
\end{equation}
where the measures are defined via: $ {\cal D}^2 \zeta_{\vec{k}}={\cal D}{\rm Re}\zeta_{\vec{k}}\ {\cal D}{\rm Im}\zeta_{\vec{k}}$ and likewise for ${\cal D}^2 h^{(\sigma)}_{\vec{k}}$. Using the wavefunctions in eq.(\ref{eq:wavefunctions}), this condition becomes

\begin{equation}
\label{eq:normalization2}
\int {\cal D}^2 \zeta_{\vec{k}}\ \prod_{\sigma=+,\times} {\cal D}^2 h^{(\sigma)}_{\vec{k}}\ \exp\left[-2 \left(\begin{array}{cc}\zeta_{-\vec{k}}, & \mathbf{h}_{-\vec{k}}\end{array}\right)\left(\begin{array}{cc}A_{k R} & \mathbf{C}_{k R}^T \\\mathbf{C}_{k R} & \mathbb{B}_{k R}\end{array}\right)\left(\begin{array}{c}\zeta_{\vec{k}} \\ \mathbf{h}_{\vec{k}}\end{array}\right)\right]< \infty,
\end{equation}
with the subscript $R$ denoting the real part, and we have taken the two polarization states and made them into a vector $\mathbf{h}_{\vec{k}}$. Normalizability requires that the (hermitian) matrix in the quadratic form inside the exponential, which we will denote by $\mathbb{M}_k$, have only {\em positive} eigenvalues, which requires both the trace and the determinant of $\mathbb{M}_k$ to be positive. Furthermore, we need to demand that in the absence of mixing, i.e. when $\mathbf{C}_{k R}=\mathbf{0}$, the state is still normalizable. These requirements then force $A_{k R}>0,\ {\rm Tr}(\mathbb{B}_{k R})>0$. The characteristic polynomial of $\mathbb{M}_k$ is
\begin{equation}
\label{eq:characteristic}
-\lambda^3 + {\rm Tr}(\mathbb{M}_k) \lambda^2 + ({\mathbf{C}_{k R}}^T {\mathbf{C}_{k R}}-A_{k R} {\rm Tr}(\mathbb{B}_{k R})-\det(\mathbb{B}_{k R}))\lambda+\det(\mathbb{M}_k)=0.
\end{equation}
Descartes rule of signs tells us that we will have three real positive roots if we have three sign changes in the coefficients of the powers of $\lambda$. Since we know that $ {\rm Tr}(\mathbb{M}_k),\ \det(\mathbb{M}_k)$ have to be positive, then we must require 

\begin{equation}
\label{eq:normconstraint}
{\mathbf{C}_{k R}}^T {\mathbf{C}_{k R}}-A_{k R} {\rm Tr}(\mathbb{B}_{k R})-\det(\mathbb{B}_{k R})<0. 
\end{equation}
Now that we know what it takes to make the state normalizable, we can actually do the functional integrals to find that the normalization condition for the state in eq.(\ref{eq:fullstate}):

\begin{equation}
\label{eq:normalizationcondition}
\left|N_k\right|^2 \frac{\pi^3}{8 \det \mathbb{M}_k}=1.
\end{equation}
We can now use this to find the various two-point functions needed for the calculation of the CMB temperature anisotropies. Consider the $\zeta$ two-point function:
\begin{equation}
\label{eq:zeta2point}
\langle \zeta_{\vec{k}} \zeta_{-\vec{k}}\rangle = \frac{\langle \psi_{\vec{k}} | \zeta_{\vec{k}} \zeta_{-\vec{k}} | \psi_k \rangle}{\langle\psi_{\vec{k}} | \psi_{\vec{k}}\rangle} = \frac{\int \mathcal{D}^2\zeta_{\vec{k}} \;\mathcal{D}^2 h^+_{\vec{k}}\; \mathcal{D}^2 h^{\times}_{\vec{k}} \;\zeta_{\vec{k}}\zeta_{-\vec{k}}\; |\psi_{\vec{k}}|^2 |\psi_{-\vec{k}}|^2}{\int \mathcal{D}^2\zeta_{\vec{k}} \;\mathcal{D}^2 h^+_{\vec{k}}\; \mathcal{D}^2 h^{\times}_{\vec{k}} \; |\psi_{\vec{k}}|^2 |\psi_{-\vec{k}}|^2}.
\end{equation}
We can obtain this as the functional derivative of log of the denominator in the above equation with respect to $A_{k R}$:

\begin{eqnarray}
&&\langle \zeta_{\vec{k}} \zeta_{-\vec{k}}\rangle = -\frac{1}{2}\frac{\partial}{\partial A_{k R}} \ln \left(\int \mathcal{D}^2\zeta_{\vec{k}} \;\mathcal{D}^2 h^+_{\vec{k}}\; \mathcal{D}^2 h^{\times}_{\vec{k}} \; |\psi_{\vec{k}}|^2 |\psi_{-\vec{k}}|^2\right) =\nonumber\\
&&= -\frac{1}{2}\frac{\partial}{\partial A_{k R}} \ln\left( \frac{\pi^3}{8 \det \mathbb{M}_k}\right) = +\frac{1}{2}\frac{\partial}{\partial A_{k R}} \ln  \det \mathbb{M}_k.
\end{eqnarray}
The determinant $ \det \mathbb{M}_k$ is easy to calculate:

\begin{equation}
\label{eq:determinant}
\det \mathbb{M}_k = A_{k R} \left(b_{0 k R}^2 -\vec{b}_{k R}^2\right)-\mathbf{C}_{k R}^T \mathbf{C}_{k R} b_{0 k R} + \mathbf{C}_{k R}^T \vec{\sigma}\cdot \vec{b}_{k R} \mathbf{C}_{k R}.
\end{equation}
From this we then find 
\begin{equation}
\label{eq:zetacorr}
\langle \zeta_{\vec{k}} \zeta_{-\vec{k}}\rangle  = \frac{1}{2} \frac{b_{0 k R}^2 -\vec{b}_{k R}^2}{A_{k R} \left(b_{0 k R}^2 -\vec{b}_{k R}^2\right)-\mathbf{C}_{k R}^T \mathbf{C}_{k R}\  b_{0 k R} + \mathbf{C}_{k R}^T\  \vec{\sigma}\cdot \vec{b}_{k R}\  \mathbf{C}_{k R}}.
\end{equation}
The other two-point functions can be found in the same way, by taking derivatives with respect to $\mathbf{C}_{k R}^{(s)}$ and $\mathbb{B}_{k S R}^{(\sigma,\sigma^{\prime})}$:

\begin{eqnarray}
\label{eq:othercorrs}
\langle \zeta_{\vec{k}} h^{(\sigma)}_{-\vec{k}}+\zeta_{-\vec{k}} h^{(\sigma)}_{\vec{k}}\rangle &=& \frac{-\mathbf{C}_{k R}^{(\sigma)} b_{0 k R} + \vec{b}_{k R}\cdot \left(\vec{\sigma}\ \mathbf{C}_{k R}\right)^{(\sigma)}}{A_{k R} \left(b_{0 k R}^2 -\vec{b}_{k R}^2\right)-\mathbf{C}_{k R}^T \mathbf{C}_{k R}\  b_{0 k R} + \mathbf{C}_{k R}^T\  \vec{\sigma}\cdot \vec{b}_{k R}\  \mathbf{C}_{k R}},\nonumber\\
\langle h^+_{\vec{k}} h^+_{-\vec{k}}\rangle &=& \frac{1}{2} \frac{A_{k R} \left(b_{0 k R}-b_{3 k R}\right)-C^{\times 2}}{A_{k R} \left(b_{0 k R}^2 -\vec{b}_{k R}^2\right)-\mathbf{C}_{k R}^T \mathbf{C}_{k R}\  b_{0 k R} + \mathbf{C}_{k R}^T\  \vec{\sigma}\cdot \vec{b}_{k R}\  \mathbf{C}_{k R}},\nonumber\\
 \langle h^+_{\vec{k}} h^{\times}_{-\vec{k}}\rangle &=& \frac{1}{2} \frac{-A_{k R} b_{1 k R}-C^+ C^{\times}}{A_{k R} \left(b_{0 k R}^2 -\vec{b}_{k R}^2\right)-\mathbf{C}_{k R}^T \mathbf{C}_{k R}\  b_{0 k R} + \mathbf{C}_{k R}^T\  \vec{\sigma}\cdot \vec{b}_{k R}\  \mathbf{C}_{k R}},\nonumber\\
\langle h^{\times}_{\vec{k}} h^{\times}_{-\vec{k}}\rangle &=& \frac{1}{2} \frac{A_{k R} \left(b_{0 k R}+b_{3 k R}\right)-C^{+  2}}{A_{k R} \left(b_{0 k R}^2 -\vec{b}_{k R}^2\right)-\mathbf{C}_{k R}^T \mathbf{C}_{k R}\  b_{0 k R} + \mathbf{C}_{k R}^T\  \vec{\sigma}\cdot \vec{b}_{k R}\  \mathbf{C}_{k R}}.
\end{eqnarray}
We can simplify these formulae somewhat by noting that 
\begin{equation}
A_{k R} = \frac{-i\alpha^2 W\left[f_k, f_k^*\right]}{2 \left|f_k\right|^2} = \frac{\alpha^2}{2 \left|f_k\right|^2},
\end{equation}
where $ W\left[f_k, f_k^*\right]$ is the Wronskian between the mode and its conjugate and we have chosen it to be $-i$ to make $A_{k R}>0$ as needed for normalization of the $\zeta$ part of the wavefunction in the absence of entanglement. Likewise, since we also require ${\rm Tr} \mathbb{B}_{k S R}>0$, we have
\begin{equation}
b_{0 k R} = \frac{\beta^2}{2 \left|g_k\right|^2}.
\end{equation}

\section{\label{sec:numerical} Numerical Results}

\subsection{\label{subsec:setup}Set Up}
We are concerned with the range of $k$-modes accessible to observations. For simplicity, and in keeping with the short inflation picture, we set the beginning of inflation to coincide with the horizon exit of the first observable mode. This does a good job of capturing the motivation of this work as discussed in the introduction. Furthermore, if too much inflation has occurred prior to the horizon exit of observable modes, excessive particle production would ensue due to the non-BD nature of our state, and the backreaction of these particles would interfere with the inflationary phase. The initial time $\tau=\tau_0$ is then defined by the horizon exit of the lowest visible wavenumber $k_{min} = -1/ \tau_0$. The final time, will be set close to the end of inflation in order to assure that all the relevant modes are well outside the horizon. 
 
We would like to compare the power spectra and the CMB temperature anisotropies we obtain from this state to those we would find in the regular unentangled case. Thus, we choose initial conditions for the pure $\zeta$ and $h^{(\sigma)}$, such that \emph{if there were no entanglement}, we would just get the standard results\footnote{Our initial state at time $\tau_0$ is not a Bunch-Davies state, we only set the \emph{initial values} of the mode functions $f_k$ and $g_k$ to their BD values to better compare with the standard result. The initial values of the entanglement parameters, $\vec{b}_k(\tau=\tau_0)$ and $\mathbf{C}_k (\tau=\tau_0)$ are non-zero and therefore our state is an excited state induced by entanglement. If the entanglement parameters were zero at $\tau_0$, the time evolution of the fields would be identical to Bunch-Davies.}. This corresponds to the Bunch-Davies vacuum states at initial time $\tau_0$ and their derivatives:
\begin{eqnarray}
\label{eq:initcondsstandard}
& & f_k(\tau=\tau_0) = \frac{\sqrt{-\pi \tau}}{2}\ H^{(1)}_{\nu_{\zeta}}(-k \tau_0),\; \;\partial_{\tau} f_k(\tau)\left|_{\tau=\tau_0}\right  . = \partial_{\tau}\left(\frac{\sqrt{-\pi \tau}}{2}\ H^{(1)}_{\nu_{\zeta}}(-k \tau)\right)\left|_{\tau=\tau_0}\right), \nonumber \\
 & & g_k(\tau=\tau_0) = \frac{\sqrt{-\pi \tau}}{2}\ H^{(1)}_{\nu_{\gamma}}(-k \tau_0),\;\;  \partial_{\tau} g_k(\tau)\left|_{\tau=\tau_0}\right . = \partial_{\tau} \left(\frac{\sqrt{-\pi \tau}}{2}\ H^{(1)}_{\nu_{\gamma}}(-k \tau)\right)\left|_{\tau=\tau_0}\right).
 \end{eqnarray}
Where $f_k(\tau=\tau_0)= f^{BD}_k$, $g_k(\tau=\tau_0) = g^{BD}_k$ and the initial values $\vec{b}_k(\tau=\tau_0),\ \mathbf{C}_k (\tau=\tau_0)$ are free parameters measuring the amount of entanglement. 

  \subsection{Bounds on Initial Entanglement Parameters}    
  Given the normalization constraints:
  \begin{eqnarray}
 && A_{kR} >0,  \quad \rm Tr (\mathbb{B}_{k R})>0\quad \rm Tr (\mathbb{M}_{k R})>0, \quad \det (\mathbb{M}_{k R})>0 ,\\
  &&{\mathbf{C}_{k R}}^T {\mathbf{C}_{k R}}-A_{k R} {\rm Tr}(\mathbb{B}_{k  R})-\det(\mathbb{B}_{k  R})<0.
  \end{eqnarray}
we can calculate the bounds on the initial magnitudes of the entanglement parameters. In terms of the magnitudes and phases of the mode functions, the above constraints are (listed in the same order):
  
 \begin{equation}
\frac{\alpha^2}{2|f_k|^2}>0, \quad \frac{\beta^2}{|g_k|^2}>0,
\end{equation}
 \begin{eqnarray}
 &&1-4\left[|\tilde{b}_{1k}|^2\cos^2(\theta_{b_1} -2\theta_g) +|\tilde{b}_{3k}|^2\cos^2(\theta_{b_3} -2\theta_g) + |\tilde{C}_k^+|^2 \cos^2(\theta_+ -\theta_f-\theta_g) \right. \nonumber \\ 
 && \quad [1- 2 |\tilde{b}_{3k}|\cos(\theta_{b_3}-2\theta_g)]  + |\tilde{C}_k^{\times}|^2 \cos^2(\theta_{\times} -\theta_f-\theta_g)[1+ 2 |\tilde{b}_{3k}|\cos(\theta_{b_3} -2\theta_g)] \nonumber \\
 &&\left.\quad-4 |\tilde{b}_{1k}|  |\tilde{C}_k^+| |\tilde{C}_k^{\times}|   \cos(\theta_{b_1} -2\theta_g)  \cos(\theta_+ -\theta_f-\theta_g)     \cos(\theta_{\times} -\theta_f-\theta_g)\right] >0,
 \end{eqnarray}
 \begin{eqnarray}
&& \frac{\alpha^2\beta^2}{2|f_k||g_k|}\left[ 1-2( |\tilde{C}_k^+|^2 \cos^2(\theta_+ -\theta_f-\theta_g)+ |\tilde{C}_k^{\times}|^2 \cos^2(\theta_{\times} -\theta_f-\theta_g))\right. \nonumber\\
&& \left. \quad+\frac{1}{2 \epsilon} \frac{|f_k|^2}{ |g_k|^2} \left[1- 4 |\tilde{b}_{31k}|^2\cos^2(\theta_{b_1}-2\theta_g) -4|\tilde{b}_{3k}|^2\cos^2(\theta_{b_3}-2\theta_g)\right] \right]>0 ,
 \end{eqnarray} 
where $\epsilon$ is the slow roll parameter and the phase angles are also $k$ dependent. We use this system of equations to find bounds on the initial magnitudes of $\tilde{b}_{1k}$,  $\tilde{b}_{3k}$,  $\tilde{C}^+_k$ and  $\tilde{C}^{\times}_k$ in terms of the magnitudes of the initial Bunch-Davies mode functions $|f_{BD}(k)|$ and $|g_{BD}(k)|$ and then use these values to evolve from.  
    
\subsection{Angular Power Spectra}
To compare our model to the CMB data we compute the angular power spectrum $C_{ll^{\prime}}$.  The angular power spectrum provides us with the spherical harmonic decomposition of the temperature anisotropies $T$, and the two components of polarization, divergence ($E$) and curl ($B$) in the CMB anisotropies. As shown above, our entangled  state modifies the evolution of the fluctuation mode functions as well as the form of the primordial power, which in their turn are used to calculate the angular power spectrum. The CMB we see today, however, does not only depend on the primordial power. After the end of inflation the universe undergoes reheating thus entering a radiation dominated phase, a process that affects the field perturbations. This is followed by recombination and a matter dominated era where the universe cools enough for photons to become free streaming, forming the CMB.  All this evolution, as well as the projection of the CMB at recombination to today is encoded in the transfer functions ($\Delta^{X}_{ls}$), where $X$ stands for $T$, $E$ or $B$ . Here we convolve our entangled primordial power spectrum with the transfer functions, calculated by the CLASS Boltzmann code \cite{Lesgourgues:2011re}, assuming the current best fit parameters released by Planck \cite{Ade:2015xua}. 

The general angular power spectra for spherical harmonic multipoles $l, l^{\prime}, m, m^{\prime}$ are defined as:
\begin{equation}
C_{ll^{\prime}mm^{\prime}}^{XX^{\prime}}  = 4 \pi \int \frac{dk}{k} \sum_{s, s^{\prime}} \left\{ \Delta^X_{ls}(k, \tau_0)\Delta^{X^{\prime}}_{l^{\prime}s^{\prime}}(k, \tau_0) \int d\Omega_{\hat{k}} P_{ss^{\prime}}(\vec{k}) \;_{-s}Y_{lm}^*(\hat{k}, \vec{e})  \;_{-s^{\prime}}Y_{l^{\prime}m^{\prime}}(\hat{k}, \vec{e})\right\},
\end{equation} 
where, $s,s^{\prime}$ are the spin weights $0, \pm 2$ to indicate scalar or tensor modes and $X, X^{\prime}$ stands for one of the possible combinations of $T$,  $E$ and $B$. The transfer function $\Delta^X_{ls}(k, \tau_0)$ for a mode $k$ at initial time $\tau_0$, follows the following parity relations for the choices of $T$,  $E$ and $B$:  

\begin{equation}
\Delta^T_{ls}(k, \tau_0) = \Delta^T_{l-s}(k, \tau_0) , \quad \Delta^E_{ls}(k, \tau_0) = \Delta^E_{l-s}(k, \tau_0), \quad \Delta^B_{ls}(k, \tau_0) = -\Delta^B_{l-s}(k, \tau_0).
\end{equation}
The two point functions for the field perturbations are related to the primordial power in the usual way:
\begin{equation}
P^{\phi, \phi^{\prime}}(k) = \frac{k^3}{2\pi^2} \langle \phi_{\vec{k}} \phi^{\prime}_{-\vec{k}}\rangle.
\end{equation}
The spin-weight primordial power $P_{ss^{\prime}}$ that appears in the $C_{ll^{\prime}}$ expression, can be written in terms of our scalar $\zeta$ fluctuation, $h^+$ and $h^{\times}$ polarization primordial power:
\begin{eqnarray}
&&P^{00}= P^{\zeta\zeta},\\
&&P^{+2+2}  = \frac{1}{2}(P^{++} + P^{\times\times} + P^{+\times}),\\
 &&P^{-2-2} =  \frac{1}{2}(P^{++} + P^{\times\times} - P^{+\times}),\\
&&P^{+2-2}  = P^{-2+2} = \frac{1}{2}(P^{++} - P^{\times\times} ),\\
&&P^{0\pm2} = \frac{1}{\sqrt{2}} (P^{0+} \pm P^{0\times}),\\
&&P^{\pm2 0} =\frac{1}{\sqrt{2}} (P^{+0} \mp P^{\times0}).
\end{eqnarray}
In the regular $\Lambda CDM$ model the angular power spectrum will only contain the terms proportional to the scalar-scalar primordial power, $P^{00}$ and the tensor-tensor primordial powers $P^{+2+2}$ and $P^{-2-2}$. All scalar-tensor power will be zero because there is no coupling (or entanglement) between the two. Moreover, the power with opposite sign spin-weights will also become zero because the the primordial power for the $+$ and $\times$ polarizations are equal in $\Lambda CDM$ and will therefore cancel \cite{Chen:2014eua, Watanabe:2010bu}.

In our model we have entanglement between tensor and scalar modes and our angular power spectrum will, in general, have nonzero scalar-tensor, as well as  the opposite sign spin-weights primordial power.  This will result in non-zero off diagonal terms ($l\neq l^{\prime}$) because we are now breaking rotational symmetry. Mathematically this is expressed by the integral over the spin weighted spherical harmonics. The $P^{+2-2} , P^{-2+2}, P^{0\pm2} , P^{\pm2 0}$ terms of the angular power will not be proportional to $\delta_{ll^{\prime}}$ for our model (they are however still proportional to $\delta_{mm^{\prime}}$).

\subsection{Primordial Power}
As mentioned above the minimal consistent choices of entanglement parameters are either $\tilde{b}_{1 k}\neq 0,\ \tilde{C}^+=\pm \tilde{C}^{\times}$ or $\tilde{b}_{3 k}\neq 0$ with one of  $\tilde{C}^+,\ \tilde{C}^{\times}$ vanishing. For computational simplicity we will therefore study these four cases in particular.  The case with $\tilde{b}_{1 k}\neq 0$ will be referred to as \emph{Case 1}, while the  $\tilde{b}_{3 k}\neq 0$ case will be \emph{Case 2}. The two point functions for these cases will then simplify considerably in terms of the mode functions $f_k(\tau)$ and $g_k(\tau)$ (Appendix 1).
The following naming scheme for the four \emph{Cases} will be used:

\begin{table}[!htp]
  \centering
  \begin{tabular}{|M{6cm}|M{6cm}|M{2cm}|}
    \hline
    \emph{Case 1} $\;(\tilde{C}^+ = \tilde{C}^{\times}) \;\equiv $ \emph{Case} $1p$ & \emph{Case 1} $\;(\tilde{C}^+ = -\tilde{C}^{\times}) \equiv \;$\emph{Case} $1m$ &\emph{Case 1}\\ [8pt] \hline
 \emph{Case 2}\;\;$(\tilde{C}^+ = 0) \;\;\;\;\; \equiv $ \emph{Case} $2\times $   &  \emph{Case 2}\;\;$(\tilde{C}^{\times} = 0) \;\;\;\;\;\;\equiv \;$\emph{Case} $2+$ &\emph{Case 2} \\ [8pt] \hline
  \end{tabular}
\end{table}

\section{Results}
\subsection{Oscillations in the Angular Power Spectrum}
\begin{figure}[ht]
\begin{center}
\begin{subfigure}[b]{0.91\textwidth}
\includegraphics[scale=0.31]{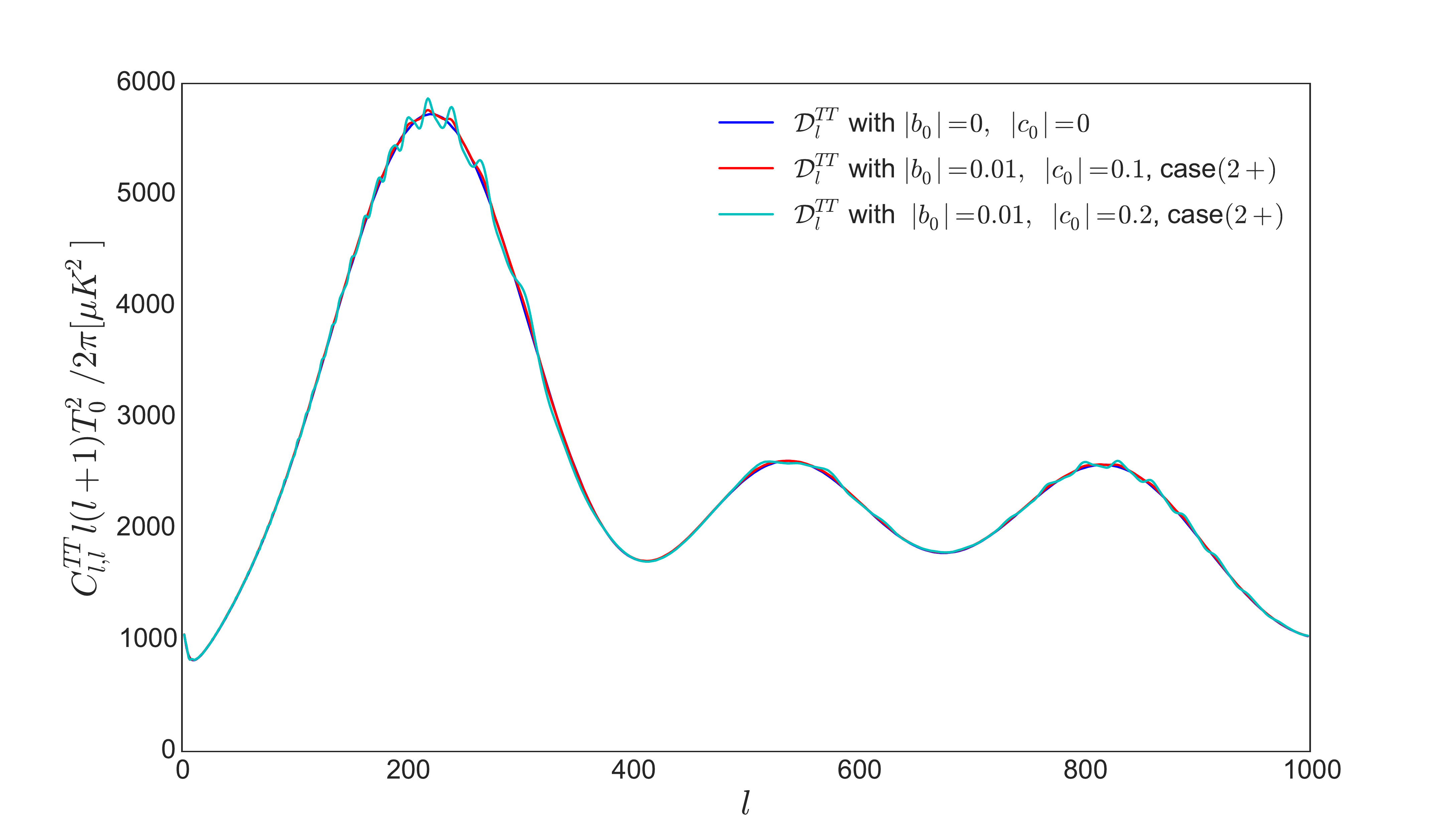}
\caption{}
\label{fig:TTH}
\end{subfigure}
\begin{subfigure}[b]{0.91\textwidth}
\includegraphics[scale=0.31]{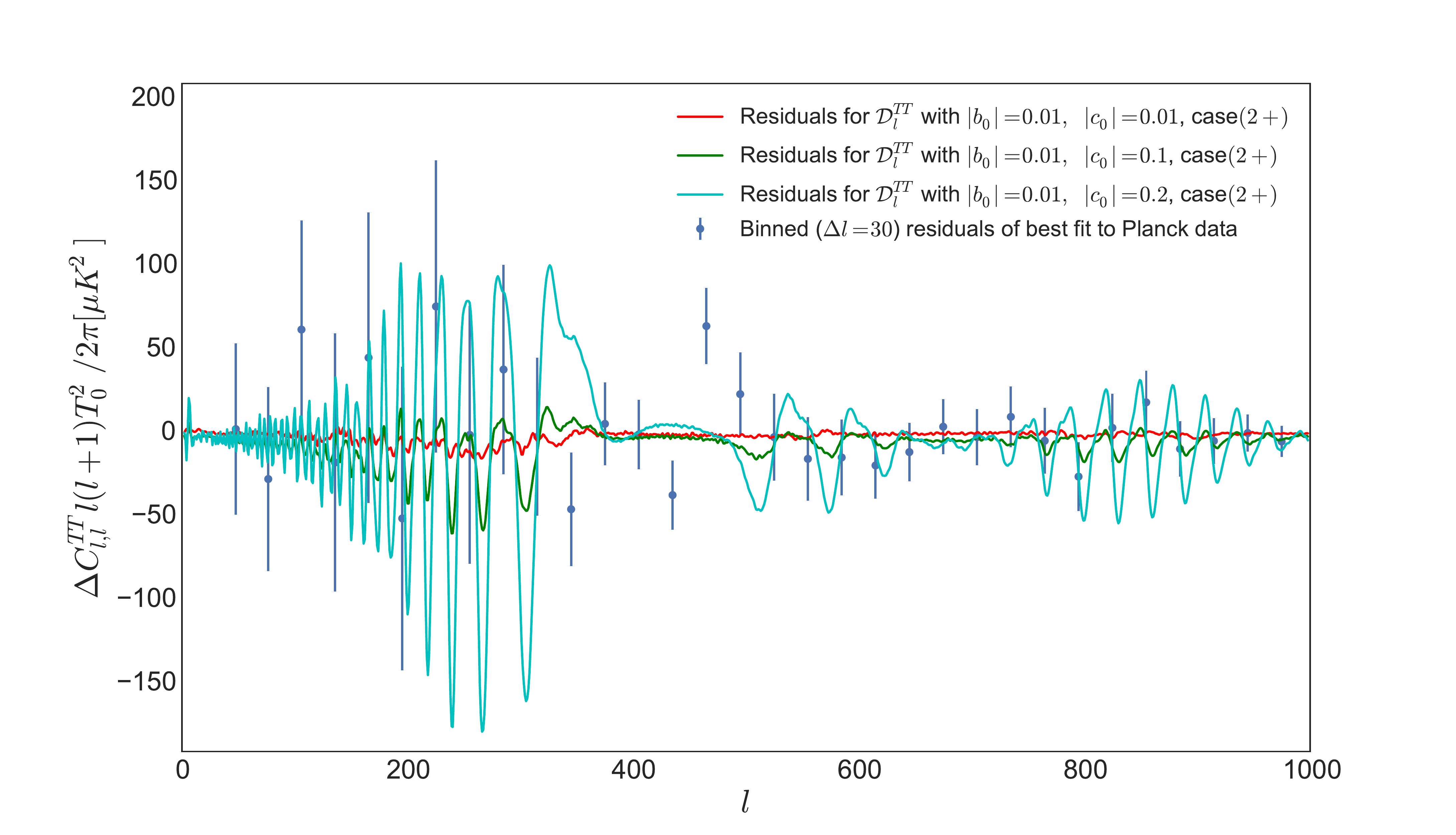}
 \caption{}
  \label{fig:residTT-case11H}
  \end{subfigure}
  \caption{(a) Temperature fluctuation angular power spectrum $C^{TT}_l$ for different values of entanglement parameter $|\tilde{C}^+_{k0}|$ ($|c_0|$ on plot to simplify labeling), keeping $|\tilde{b}_{k 3}(\tau_0)|$ ($|b_0|$ on plot) constant for \emph{Case 2+}, compared to the zero-entanglement angular power. (b) Difference between the zero entanglement $C_l^{TT}$ and non-zero entanglement $C_l^{TT}$ for different values of entanglement parameter $|\tilde{C}^+_{k0}|$ \emph{Case 2+} with the residual binned Planck data. In both figures, the larger oscillations correspond to larger values of   $|\tilde{C}^+_{k0}|$.} 
  \end{center}
\end{figure}

\begin{figure}[ht]
\begin{center}
\begin{subfigure}[b]{0.91\textwidth}
\includegraphics[scale=0.31]{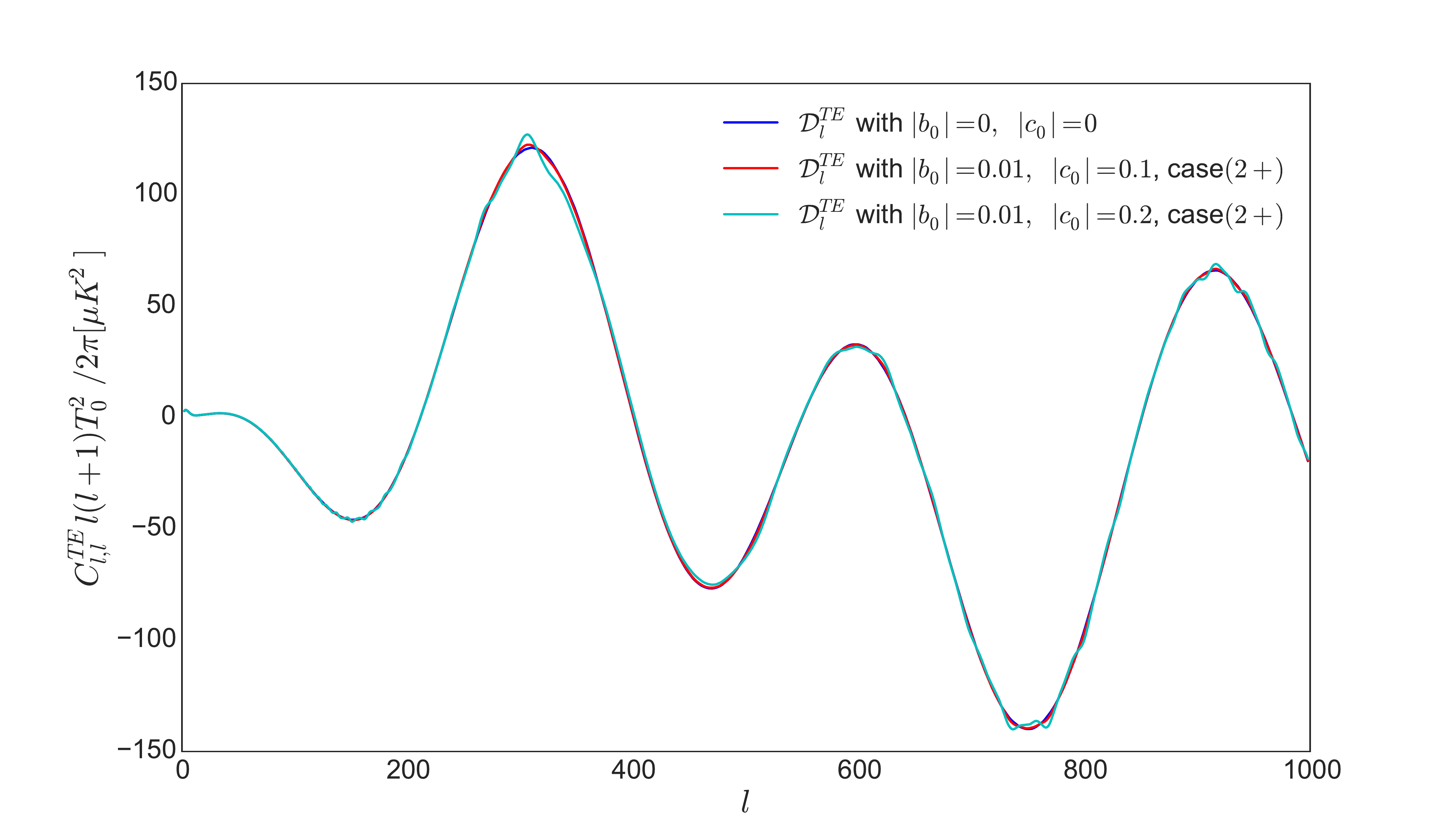}
\caption{}
  \label{fig:TE-case11}
\end{subfigure}
\begin{subfigure}[b]{0.91\textwidth}
\includegraphics[scale=0.31]{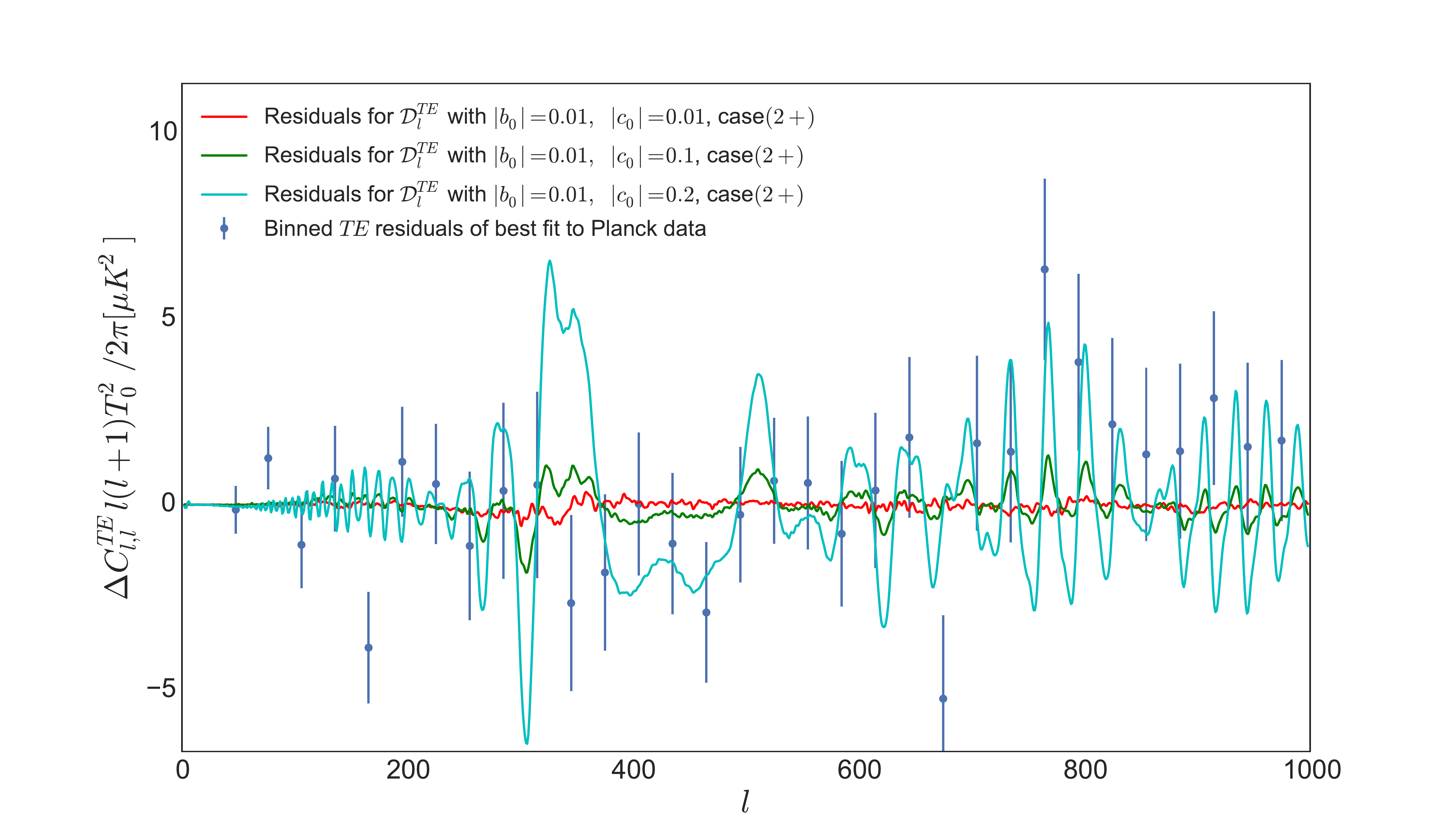}
 \caption{}
    \label{fig:residTE-case11H}
  \end{subfigure}
  \caption{(a) Polarization angular power spectrum $C^{TE}_l$ for different values of entanglement parameter $|\tilde{C}^+_{k0}|$ ($|c_0|$ on plot), keeping $|\tilde{b}_{k 3}(\tau_0)|$ ($|b_0|$ on plot) constant for \emph{Case 2+}, compared to the zero-entanglement angular power. (b) Difference between the zero entanglement $C_l^{TE}$ and non-zero entanglement $C_l^{TE}$ for different values of entanglement parameter $|\tilde{C}^+_{k0}|$ \emph{Case 2+} with the residual binned Planck data. In both figures, the larger oscillations correspond to larger values of   $|\tilde{C}^+_{k0}|$.} 
  \end{center}
\end{figure}

\begin{figure}[ht]
\begin{center}
\begin{subfigure}[b]{0.91\textwidth}
\includegraphics[scale=0.31]{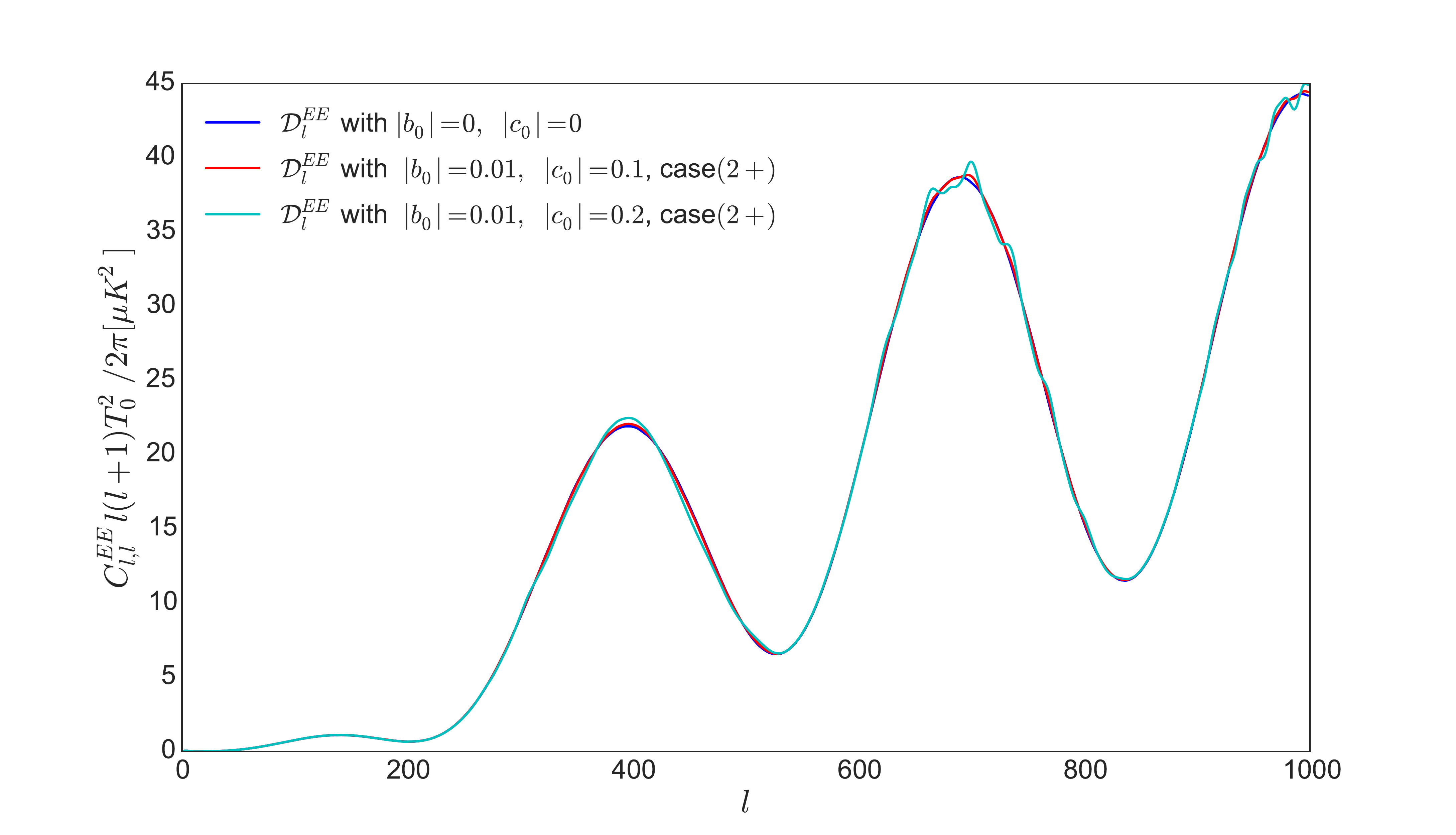}
\caption{}
  \label{fig:EE-case11}
\end{subfigure}
\begin{subfigure}[b]{0.91\textwidth}
\includegraphics[scale=0.31]{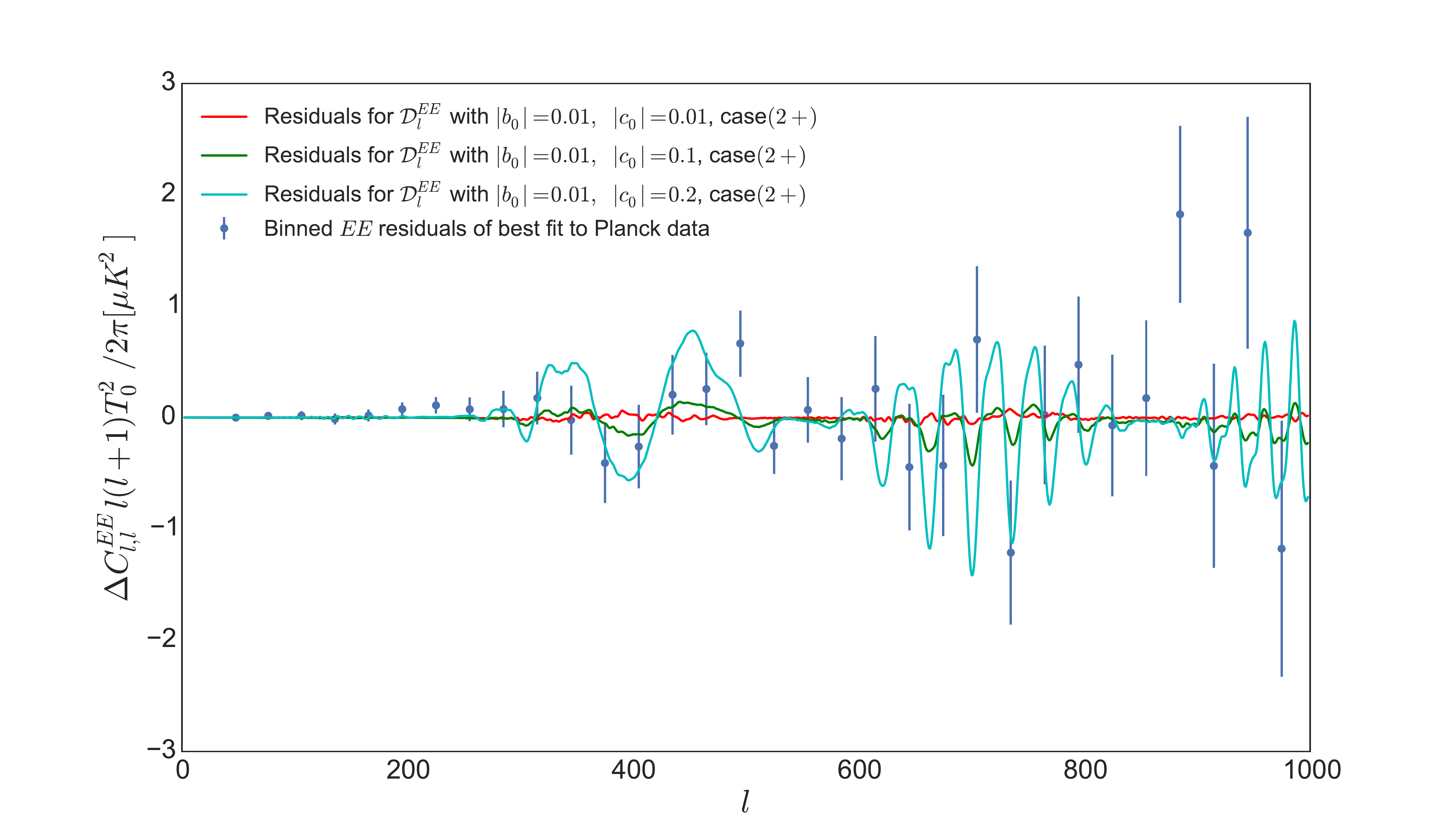}
 \caption{}
    \label{fig:residEE-case11H}
  \end{subfigure}
  \caption{(a) Polarization angular power spectrum $C^{EE}_l$ for different values of entanglement parameter $|\tilde{C}^+_{k0}|$ ($|c_0|$ on plot), keeping $|\tilde{b}_{k 3}(\tau_0)|$ ($|b_0|$ on plot)  constant for \emph{Case 2+}, compared to the zero-entanglement angular power. (b) Difference between the zero entanglement $C_l^{EE}$ and non-zero entanglement $C_l^{EE}$ for different values of entanglement parameter $|\tilde{C}^+_{k0}|$ \emph{Case 2+} with the residual binned Planck data. In both figures, the larger oscillations correspond to larger values of   $|\tilde{C}^+_{k0}|$.} 
  \end{center}
\end{figure}

\begin{figure}[ht]
\centering
\includegraphics[scale=0.30]{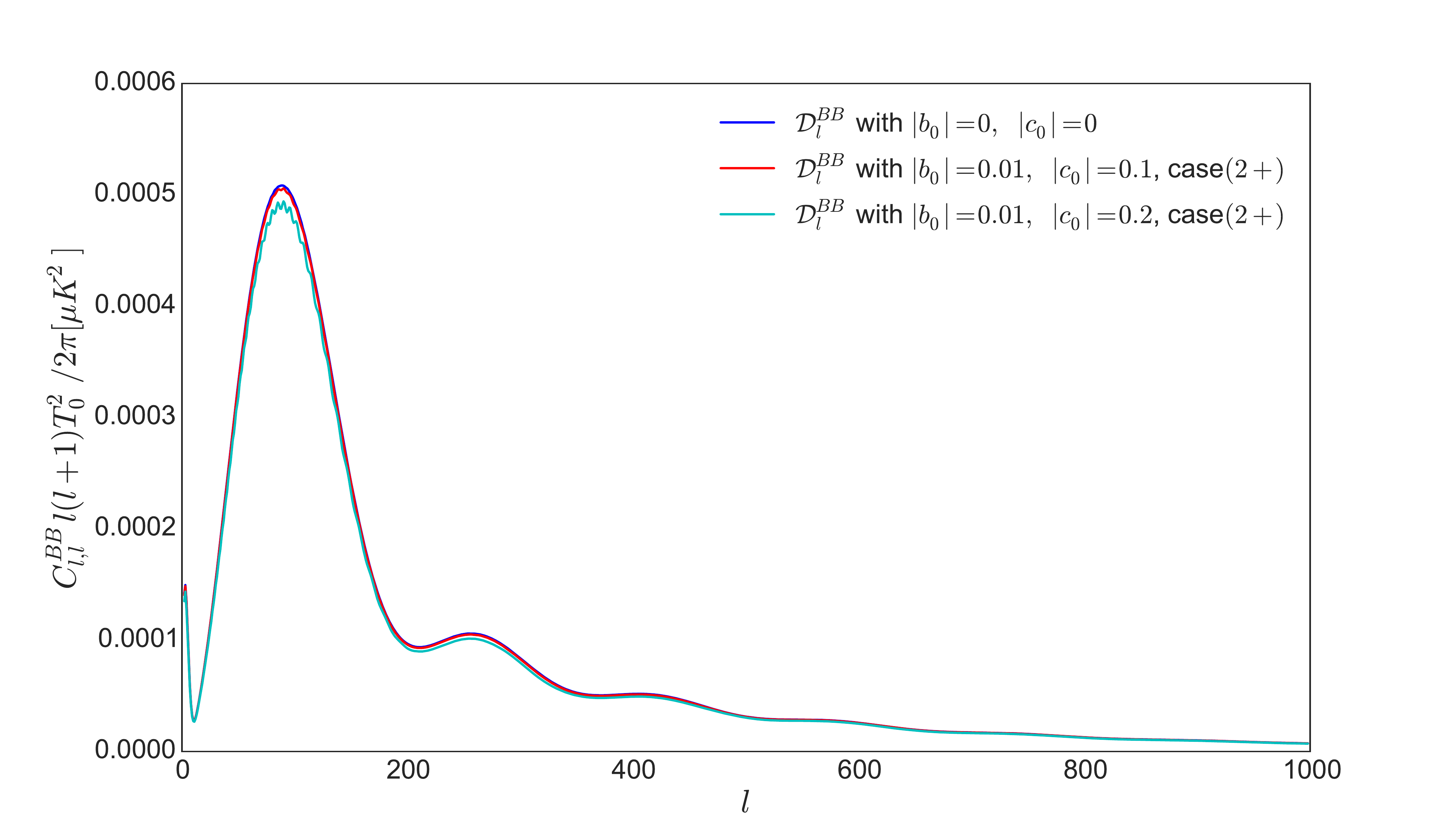}
 \caption{Polarization angular power spectrum $C^{BB}_l$ for different values of entanglement parameter $|\tilde{C}^+_{k0}|$ ($|c_0|$ on plot), keeping $|\tilde{b}_{k 3}(\tau_0)|$ constant for \emph{Case 2+}. The larger the value of $|\tilde{C}^+_{k0}|$ the larger the amplitudes of oscillation. There is also an over all degrease in power for larger values of $|\tilde{C}^+_{k0}|$.}
  \label{fig:BB-case11}
\includegraphics[scale=0.30]{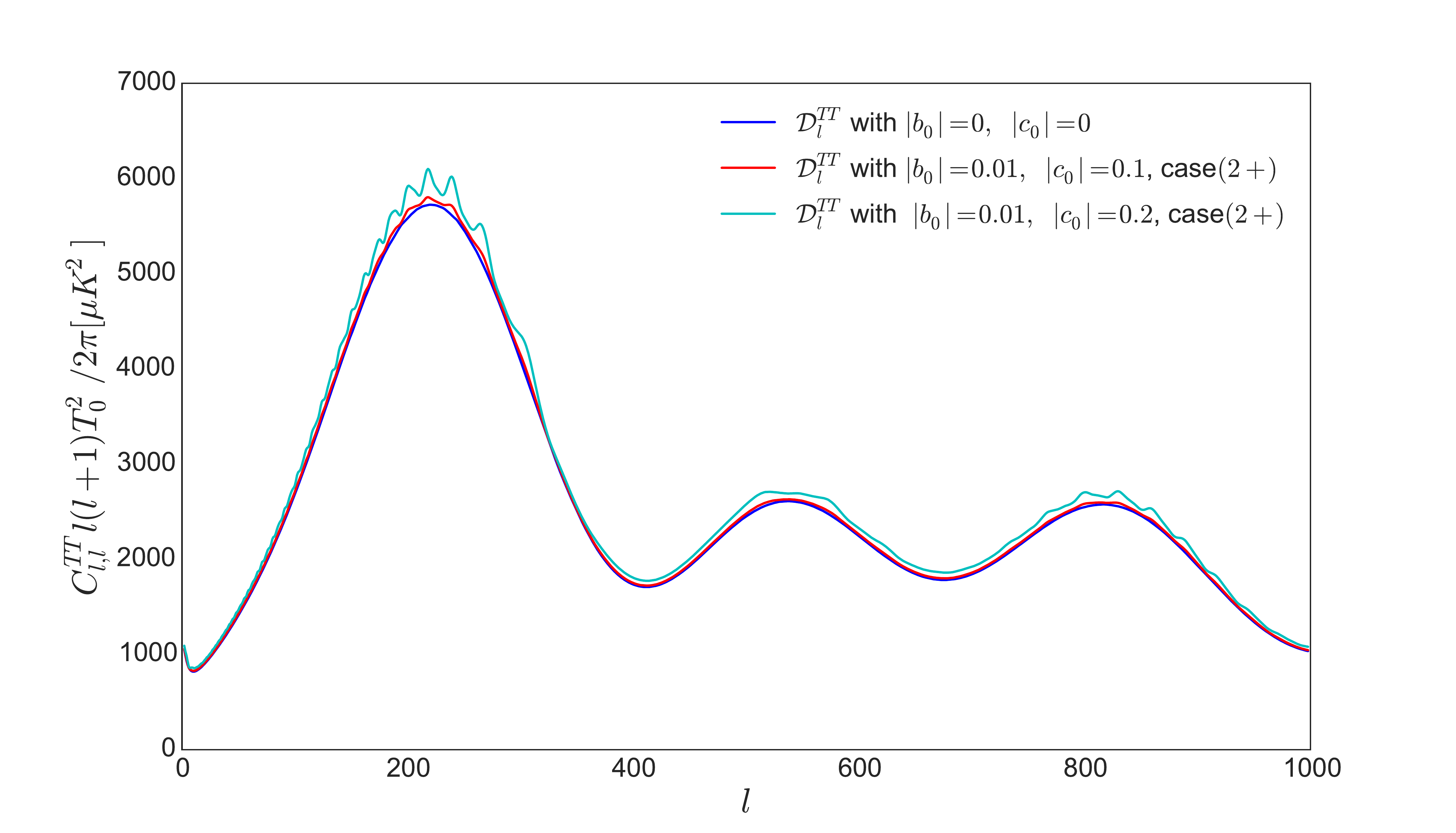}
\caption{$C^{TT}_l$ for different values of entanglement parameter $|\tilde{C}^+_{k0}|$ ($|c_0|$ on plot), keeping $|\tilde{b}_{k 3}(\tau_0)|$ ($|b_0|$ on plot) constant for \emph{Case 2+}, compared to the zero-entanglement angular power. The Hubble parameter at the onset of inflation, $H_{I0}$ is fixed for all curves in this figure. For larger $|\tilde{C}^+_{k0}|$ there is an overall increase in power as well as larger the amplitudes of oscillation. } 
\label{fig:TT}
\end{figure}

In this section we characterize what effects the non-zero entanglement in our state has on the CMB.  The most apparent signatures are small oscillations in the primordial power and the angular power spectra.  These originate from the presence of the  $k$ dependent phases of the mode functions in the two point functions, as discussed in ref.\cite{Albrecht:2014aga}. Of course, as the magnitudes of the initial values of $\tilde{b}_{k}$ and  $\tilde{C}_{k}$ ($|\tilde{b}_{k0}|,|\tilde{C}_{k0}|$) are taken to zero this effect disappears, giving us the usual $\Lambda CDM$ scenario. Due to the computationally intensive process needed to solve the non-linear coupled mode equations we leave a full MCMC analysis for a later project. In this paper we restrict ourselves  to varying the initial values of the entanglement parameters, $\tilde{b}_{k0}$ and  $\tilde{C}_{k0}$ and observing the induced changes in the angular power spectra. This gives us a feeling for how a full MCMC analysis could be used to constrain these parameters. The oscillations induced by entanglement can be clearly seen in both temperature angular power spectrum for $l=l^{\prime}$ (fig.(\ref{fig:TTH}))\footnote{ Due to the fact that the integrals over the weighted spherical harmonics get increasingly computationally intensive at at higher $l$'s we only show plots up to $l=1000$. All the power spectra here are averaged over all $m$.} as well as the $TE, EE$ and $BB$ polarization power (figs.(\ref{fig:TE-case11}, \ref{fig:EE-case11}, \ref{fig:BB-case11}), respectively). 
To give a clearer picture of how much these $C_l$'s differ from those of the $\Lambda CDM$ model, we plot the difference of the zero-entanglement best fit $C_l$ and our model's $C_l$, with non-zero entanglement, on top of the binned residual data given by Planck (figs.(\ref{fig:residTT-case11H}, \ref{fig:residTE-case11H}, \ref{fig:residEE-case11H}))\footnote{ Since our calculations do not include lensing effects the residual data we use has been obtained by subtracting the lensed best fit power. The ``residual" line for our model is, on the other hand, obtained by subtracting the non-zero entanglement $C_l$ from the the non-lensed best fit. We believe the effect of the lensing, caused by the `new' entanglement component of the power (i.e. the small oscillations and the tensor-scalar cross terms) will be small in comparison to the overall effect of the entanglement. We use this approximation to give the qualitative analysis we present here, while acknowledging that a full lensing analysis will be necessary for a systematic comparison with the data.}. We see that as the $|\tilde{C}_{k0}|$ parameters increase so do the amplitudes of oscillation. For a fixed scale of inflation $H_{I}$there is also an increase in overall amplitude of the $C_l$'s (fig.(\ref{fig:TT})). However, in our model the scale of inflation is also a free parameter so it can be adjusted for each set of entanglement parameters to rescale the $C_l$ to match the best fit more closely, (figs.(\ref{fig:residTT-case11H}, \ref{fig:residTE-case11H}, \ref{fig:residEE-case11H}))\footnote{ Each primordial power is scaled by a factor of $(H_I/H_{pl})^2$. The scale of inflation $H_I$ would be one of the free parameters which would be varied over when doing a MCMC analysis. Having not yet done this, the plots we show here (with exception of fig.(\ref{fig:TT}) which has the same $H_I$ for all $C_l$'s) are an estimation of what the rescaled $C_l$'s with different scales of inflation would look like when finding the best fit.}. Clearly, the amplitude of oscillations present one way of constraining the parameter $|\tilde{C}_{k0}|$, and hence can tell us how much, if any, entanglement between scalar and tensor modes can exist at the beginning of inflation given our current data.
For plots of the $C_l$'s for the different \emph{Cases} see Appendix 2. In both temperature and polarization power spectra, for a given set of entanglement parameters,  \emph{Case \;1p} and \emph{Case \;1m} ($\tilde{C}^+ = \pm\tilde{C}^{\times}$)  exhibits larger oscillation amplitudes then  \emph{Case \;$2+$} and \emph{Case \;$2\times$}.

\clearpage
 \subsection{The Second Entanglement Parameter}  
Increasing the initial $|\tilde{b}_{k}|$, while holding $|\tilde{C}_{k}|$ fixed, causes a less drastic effect then increasing the initial $|\tilde{C}_{k}|$. Our numerical explorations of larger initial $|\tilde{b}_{k}|$ parameter space were stymied by numerical instabilities encountered when calculating the mode functions. Resolving these  would require a considerable increase in precision and hence in run-time, and therefore we chose to omit these from this work. Within the range of initial $|\tilde{b}_{k}|$ that behaved well,  we saw no  significant changes in the low-$l$ behavior of the power spectrum. However since we do break rotational invariance in the presence of  non-zero $\tilde{b}_{3k}$ or $\tilde{b}_{1k}$, some alternate test of isotropy might reveal a signature akin to the large scale anomaly present in the $CMB$ data\footnote{We saw tantalizing hints of this behavior in our exploration but due to numerical issues we report on these here only to motivate a more rigorous future analysis.}. It may also be possible that an exploration of higher initial  $|\tilde{b}_{k}|$ would reveal different behavior.   

\begin{figure}[ht]
\begin{center}
\begin{subfigure}[b]{0.91\textwidth}
\includegraphics[scale=0.31]{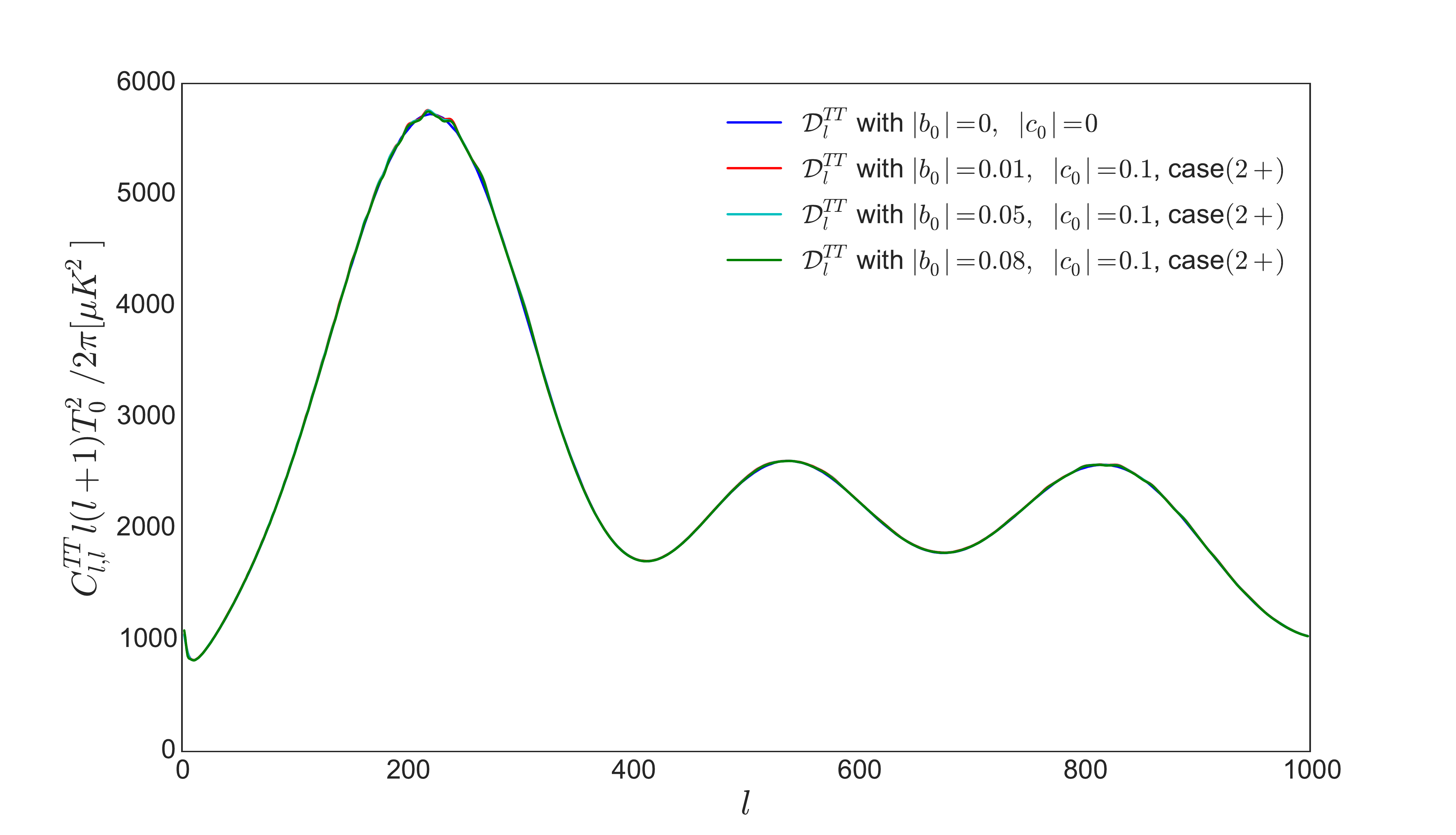}
\caption{}
    \label{fig:Tbcase}
\end{subfigure}
\begin{subfigure}[b]{0.91\textwidth}
\includegraphics[scale=0.31]{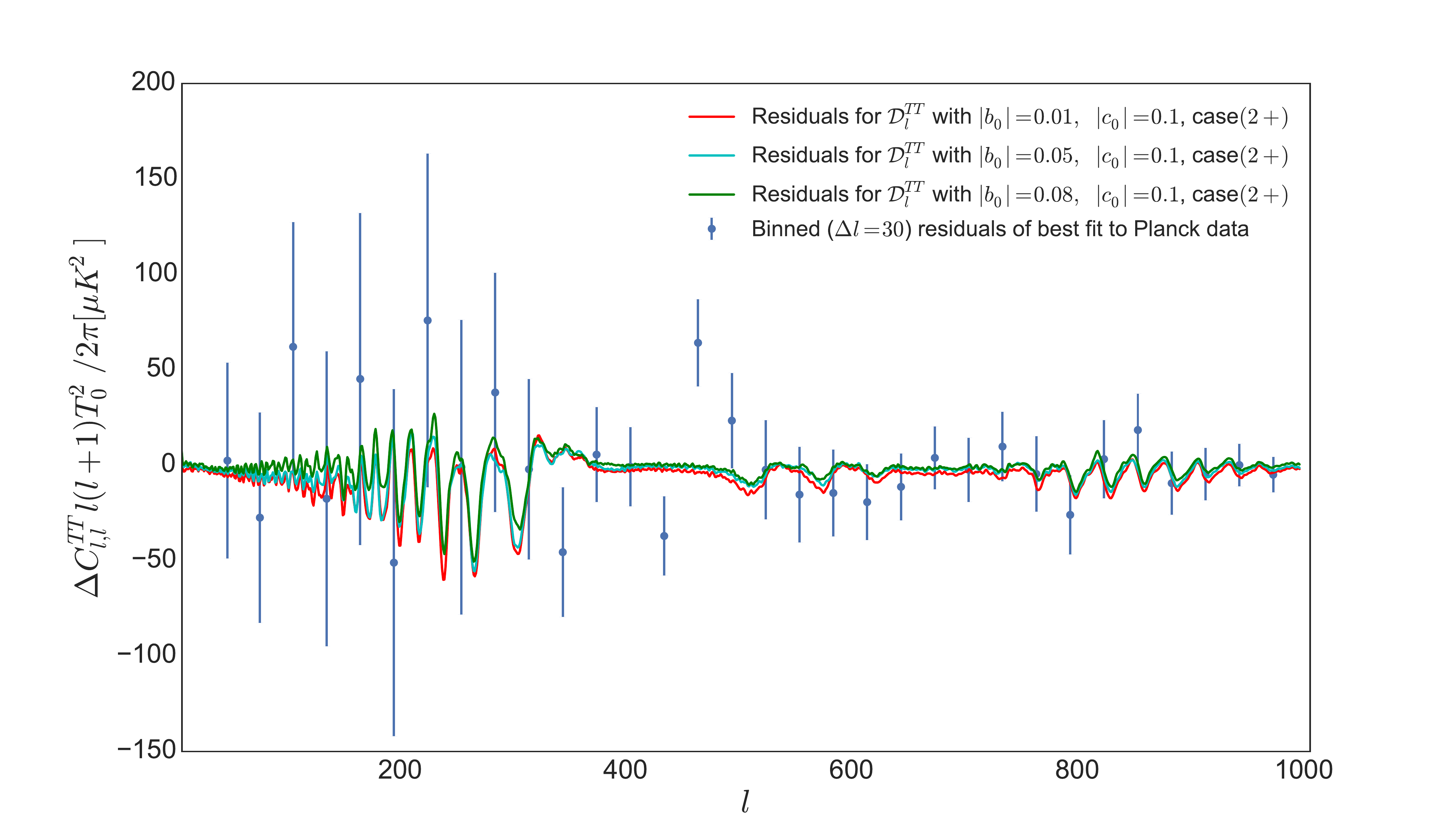}
 \caption{}
  \end{subfigure}
  \caption{(a) Temperature fluctuation angular power spectrum $C^{TT}_l$ for different values of entanglement parameter $|\tilde{b}_{k 3}(\tau_0)|$ ($|b_0|$ on plot), keeping $|\tilde{C}^+_{k0}|$ ($|c_0|$ on plot) constant for \emph{Case 2+}, compared to the zero-entanglement angular power (b) Difference between the zero entanglement $C_l^{TT}$ and non-zero entanglement $C_l^{TT}$ for different values of entanglement parameter $|\tilde{b}_{k 3}(\tau_0)|$, keeping $|\tilde{C}^+_{k0}|$ constant for \emph{Case 2+}.}
    \end{center}
\end{figure}

\subsection{The TB and EB Polarizations}
\begin{figure}[ht]
\begin{center}
\begin{subfigure}[b]{0.91\textwidth}
\includegraphics[scale=0.31]{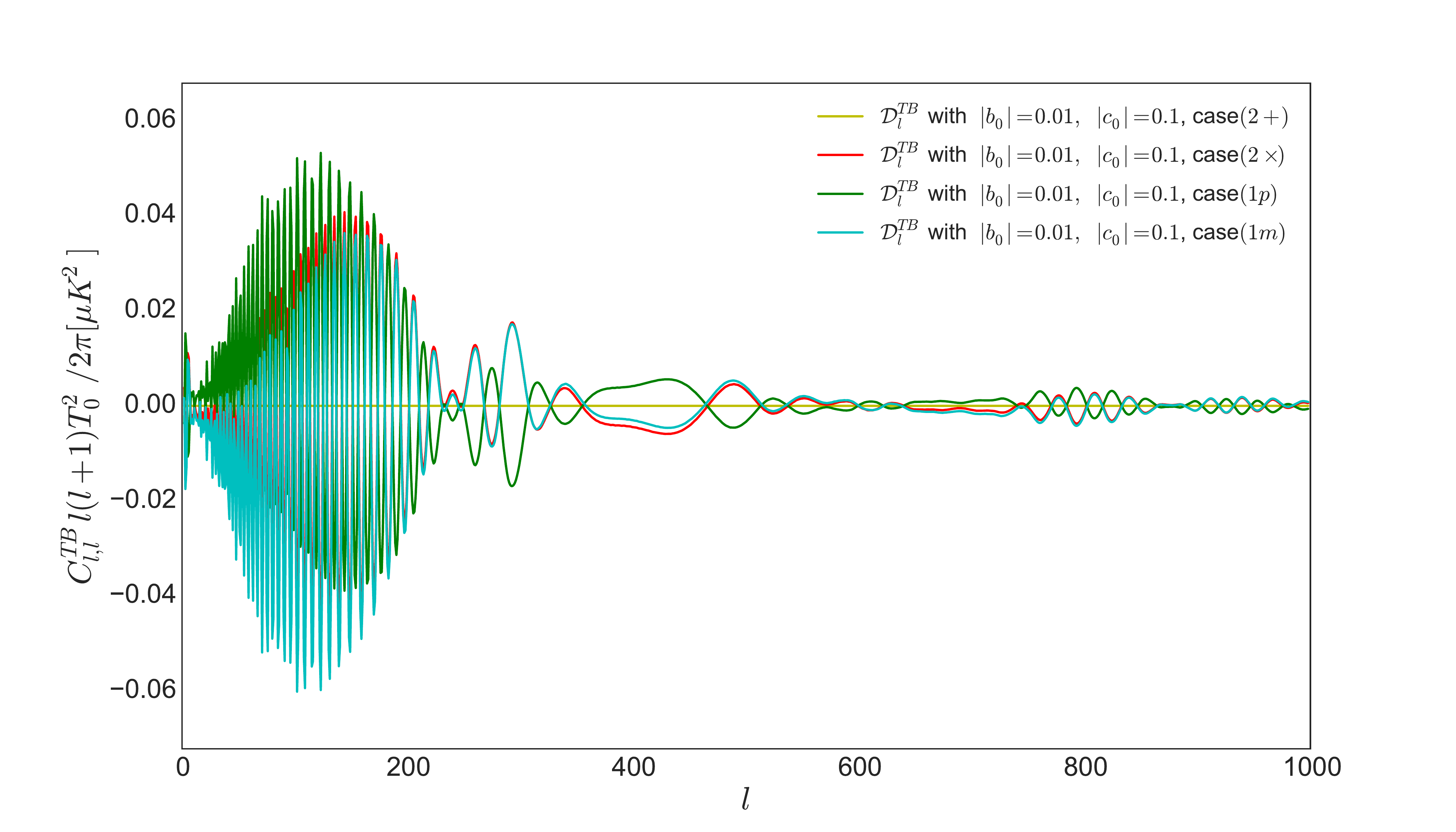}
\caption{}
 \label{fig:tb}
\end{subfigure}
\begin{subfigure}[b]{0.91\textwidth}
\includegraphics[scale=0.31]{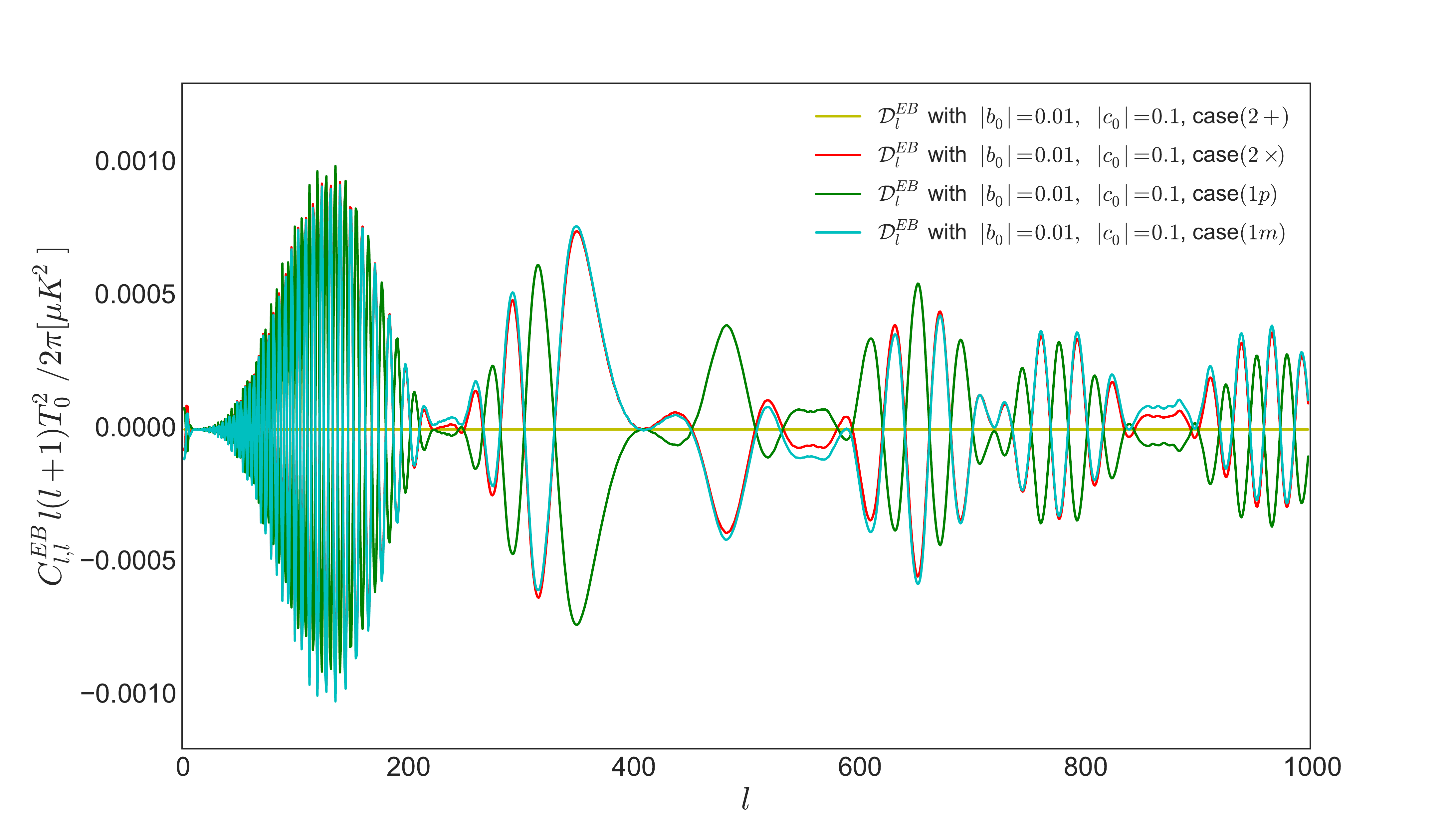}
 \caption{}
   \label{fig:eb}
  \end{subfigure}
  \caption{(a) $TB$ polarization spectra of the different \emph{Cases} for entanglement parameters $|\tilde{C}^+_{k0}| = 0.1$ ($|c_0|$ on plot) and $|\tilde{b}_{k0}| = 0.01$ ($|b_0|$ on plot). All but \emph{Case $2+$} have non zero $TB$ polarization spectra which differs from the $\Lambda CDM$ model. (b) $EB$ polarization spectra of the different \emph{Cases}  for entanglement parameters $|\tilde{C}^+_{k0}| = 0.1$ and $|\tilde{b}_{k0}| = 0.01$. All but \emph{Case $2+$} have non zero $EB$ polarization spectra which differs from the $\Lambda CDM$ model.}
  \end{center}
\end{figure}

Another important feature of our model that distinguishes it from $\Lambda CDM$ is that it has non zero contributions to the $TB$ and $EB$ polarizations. These depend solely on the cross scalar-tensor and cross-polarization two point functions:
\begin{eqnarray*}
&&\text{For $l$ and $l^{\prime}$ both even or both odd:}\\
&&C_{ll^{\prime}}^{TB, EB}  = 4 \pi \int \frac{dk}{k} \left\{  -\sqrt{2}\Delta^{T,E}_{l0}(k)\Delta^{B}_{l^{\prime}2}(k) P^{0\times}(k) \mathcal{I}^{02}_{ll^{\prime}} + \Delta^{T,E}_{l2}(k)\Delta^{B}_{l^{\prime}2}(k) P^{+ \times}(k) \delta_{ll^{\prime}} \right\} ,\\\\
&&\text{For either $l$ or $l^{\prime}$ being even and the other odd:}\\
&&C_{ll^{\prime}}^{TB, EB}  = 4 \pi \int \frac{dk}{k} \left\{  -\sqrt{2}\Delta^{T,E}_{l0}(k)\Delta^{B}_{l^{\prime}2}(k) P^{0+}(k) \mathcal{I}^{02}_{ll^{\prime}}  \right\}, \\
\end{eqnarray*}
where $\mathcal{I}^{02}_{ll^{\prime}}$ is the integral of the scalar and weighted spherical harmonics.  Note that the \emph{Case \;1} power has non-zero $\langle h^+ h^{\times} \rangle$ contributions to their primordial power, giving it an extra contribution to the $TB$ and $EB$ correlations, while, \emph{Case 2} power has no cross polarization terms. We plot an example of what such $TB$ and $EB$ correlations would look like for each of the four \emph{Cases} (figs.(\ref{fig:tb}, \ref{fig:eb})). Increasing the entanglement parameters would again increase the amplitude of the oscillations present . Presently, only the low $l$ data for these polarizations have been released, but future analysis on the higher $l$ multipoles will again provide either evidence for, or constraints on this model. The one case, (considered so far), that would be indistinguishable from $\Lambda CDM$, by looking solely at $TB$ and $EB$, for either $l$ and $l^\prime$ both being even or both being odd, would be \emph{Case $2+$} (figs.(\ref{fig:tb}, \ref{fig:eb})). \emph{Case 2+} not only has a zero $\langle h^+ h^{\times} \rangle$ two point function but also a vanishing $P^{0\times}(k)$ primordial power (see Appendix A). For either $l$ or $l^{\prime}$ being even and the other odd, however, it would have non-zero $TB$ and $EB$ amplitude while \emph{Case $2\times$} would be vanishing for this set of $l, l^{\prime}$.
  
\clearpage
\section{The Origin of the Oscillations}
The entanglement induced oscillations in the power spectra came about due to phases of our mode functions that arise in the reduced density matrix of our state\footnote{For a discussion of the origin of the oscillations with the Heisenberg picture see Appendix 4}. If we were able to make a measurement of the whole state at once the expectation values of the corresponding observables would be calculated using the whole pure density matrix $\rho = |\Psi\rangle \langle \Psi|$ in the usual way, $\langle \mathcal{O} \rangle = \rm Tr(\hat{\mathcal{O}} \rho)$. The physical observables we do measure, for example the two point function of one of the fields, however are not of the whole state but of a subset of degrees of freedom. To compute these observables we therefore, necessarily need to trace out over the `non-observed' degrees of freedom of the pure state yielding the reduced density matrix of a mixed state. To illustrate this better we resort to a toy model: a correlated harmonic oscillators in an entangled Gaussian state akin to the one used in this paper.
The entangled state is:
\begin{equation} 
\langle x, y | \Psi\rangle = \Psi(x,y,t) =  N(t) e^{-\frac{1}{2}(m_x \omega_x A(t) x^2 + m_y \omega_y B(t) y^2 + 2 \sqrt{m_x m_y} \sqrt{\omega_x \omega_y} C(t) x y)} ,
 \end{equation}
 with the SHO Hamiltonian:
 \begin{equation}
 H = \frac{p_x^2}{2 m_x} + \frac{p_y^2}{2 m_y} + \frac{1}{2} m_x \omega_x^2 x^2 + \frac{1}{2} m_y \omega_y^2 y^2 .
 \end{equation}
Integrating out degrees of freedom of the pure density matrix to produce the reduced density matrix introduces phase information which are responsible for the oscillatory behavior seen in the observables. In order to make explicit the mode functions which describe the time dependence of the width of the Gaussian in the $x$ and $y$ direction we make the following change of variables:
\begin{equation} 
i A(t) = \frac{\dot{\chi}}{\chi}, \quad \quad i B(t) = \frac{\dot{\psi}}{\psi}, \quad \quad C(t) = \frac{\lambda}{\chi \psi}.
\end{equation}
Where entanglement constant $\lambda$ modulates the amount of entanglement. The reduced density matrix for one of the variables needed to calculate observables takes the form:
\begin{equation}
\rho_{\text{red}} = \int dy \langle x, y | \Psi\rangle \langle \Psi | x^{\prime}, y\rangle = \sqrt{\frac{m_x}{\pi}} \sqrt{ A_R - \frac{C^2_R}{B_R}}\; e^{-\frac{1}{2}(\gamma x^2 + \gamma^* x^{\prime 2} - 2 \beta x x^{\prime})},
\end{equation} 
where
\begin{equation}
\gamma = A -\frac{C^2}{2B_R}, \quad\quad \beta = \frac{|C|^2}{2B_R}.
\end{equation}
The oscillatory behavior induced by the entanglement between the two coordinates is parametrized by the phase information ($\theta_{\chi}, \theta_{\psi}$), defined by $\chi = |\chi| e^{i \theta_{\chi}}$ and $\psi = |\psi| e^{i \theta_{\psi}}$,  explicitly present in the reduced density matrix and observables such as $n$-point functions. In terms of the mode functions $\chi$ and $\psi$ an $n$-point function of $x$ is,
\begin{equation}
\langle x^n \rangle \propto \frac{|\chi|^{\frac{n}{2}}}{\left(1- 4 \lambda^2 \cos^2(\theta_{\chi} +\theta_{\psi})\right)^{\frac{n}{2}}}. 
\end{equation}
In particular the mode function phase information in the $n$-point functions comes from taking the real part of the entanglement parameter C and as the entanglement constant $\lambda$ approaches zero the amplitude of the oscillations (of the cosine) will also vanish. The entanglement can also be quantifies by calculating the Von Neumann entanglement entropy:
\begin{eqnarray} 
S_{\text{ent}} &=& -\text{Tr}(\rho_{\text{red}} \ln\rho_{\text{red}})\\
&=& -\ln(1-\xi) - \frac{\xi}{1-\xi} \ln(\xi),
\end{eqnarray}
with 
\begin{equation}
\xi = \frac{\beta}{\gamma_R + \sqrt{\gamma_R^2 -\beta^2}}.
\end{equation}
The entanglement entropy (see Appendix 3 for derivation) vanishes as as $\lambda \rightarrow 0$: $$\lim_{\lambda \to\ 0} S_{\text{ent}} = 0.$$

\section{Discussion}
In the standard cosmological picture, inflaton quantum fluctuations are taken to start in the de Sitter invariant Bunch-Davies state. However, if the beginning of inflation was marked by a more complicated, yet unknown, process (such as bubble tunneling, for example \cite{Phillips:2014yma},  \cite{Albrecht:2014eaa},  \cite{Ulvestad:2012hm}, \cite{Albrecht:2011yg}, \cite{Albrecht:2009vr}, \cite{Albrecht:2004ke}) it it is possible that field modes present at that time could be in an entangled state. In this paper we tested this possibility by asking what, if any, observable effects might become imprinted on CMB observables if scalar and tensor fluctuations were entangled. It is worth noting that while we chose to entangle scalar and tensor fluctuations, the same analysis can be repeated if we entangle the scalar (or tensor or both) fluctuations with another field \cite{Albrecht:2014aga}. 

An interesting point made in ref.\cite{Collins:2016ahj} concerns the issue of whether processes such as reheating could affect the evolution of the fluctuations while they are outside the horizon. For the standard Bunch-Davies state, this possibility was ruled out by Weinberg in his discussion of adiabatic modes \cite{Weinberg:2003sw}. The situation dealt with in ref.\cite{Collins:2016ahj} in which only the {\em{initial}} state was modified, but then followed the standard evolution equations was also protected by Weinberg's analysis. It is not clear to us at this point whether this analysis applies to our state, though we should note that it is not as if we have added new operators to the Einstein action. The fact that effects from our state survive to late time gives us confidence that a variant of Weinberg's results hold in our case, but we are exploring this further. One might imagine that the lack of isotropy in the state would be incompatible with an FRW treatment of the background geometry. However, to the extent that we are keeping the back-reaction of this state on the geometry perturbatevley small we expect that our treatment will be consistent.

Our analysis revealed a number of novel and interesting features. In particular, we saw oscillations in the primordial power spectra that could survive the convolution with the transfer functions to imprint themselves in the observed angular power spectra of the CMB. These devolve from the phases in the fluctuations are present essentially due to the fact that power spectra are observables corresponding to a subset of the total degrees of freedom in the system.The amplitude of these oscillations, if observed, can therefore be used to constrain the entanglement parameters. Moreover, because scalar perturbations are entangled with the tensor ones, our model also allows for non-zero $TB$ and $EB$ correlations which would clearly distinguish our model from $\Lambda CDM$ if signals were to be observed. Finally the parameters $\tilde{b}_{1k}$ and  $\tilde{b}_{3k}$ break rotational invariance, and might be useful in the understanding of the large scale anomalies in the CMB. The small oscillations induced by entanglement could be observed and a full MCMC analysis of our model may reveal a better fit to the data then the standard $\Lambda CDM$ scenario. 

\textbf{Acknowledgements}: 
We would like to thank Marina Magliaccio, Sugumi Kanno and Jiro Soda for many helpful discussions and Yi Wang for providing us with a modified version of the CLASS code used to extract the transfer functions. We would also like to thank Tereza Vardanyan and Hael Collins for useful discussions. R. H. was supported in part by the Department of Energy under grant DE-FG03-91-ER40682. He would also like to thank the Physics Department at UC Davis for hospitality while this work was in progress. A. A. and N. B. were supported in part by DOE Grants DE-FG02-91ER40674 and DE-FG03- 91ER40674.

\clearpage

\appendix

\section{Two-Point Functions in Terms of Mode Functions}

We collect here the expressions for the primordial power spectra for all the cases we have examined. The denominator for each of the cases are:
\begingroup\makeatletter\def\f@size{9}\check@mathfonts

\subsubsection{Scalar-Scalar}
\begin{eqnarray*}
&&Case \;\;1 : \\
&&\langle \zeta_{\vec{k}} \zeta_{-\vec{k}} \rangle = \\
&&\frac{|f_k|^2}{\alpha^2} \frac{1-4 |\tilde{b}_{1k}|^2 \cos^2(\theta_{k b_1} - 2 \theta_{k g})}{1-4\left[|\tilde{b}_{1k}|^2 \cos^2(\theta_{k b_1} - 2 \theta_{kg})+ 2 |\tilde{C}|^2 \cos^2(\theta_{kC} - \theta_{kf} -\theta_{kg}) [1+ (-1)^M\; 2\; |\tilde{b}_{1k}|^2 \cos^2(\theta_{kb_1} - 2 \theta_{kg})]\right]},\\\\
&&Case \;\;2 : \\
&&\langle \zeta_{\vec{k}} \zeta_{-\vec{k}} \rangle = \\
&&\frac{|f_k|^2}{\alpha^2} \frac{1-4 |\tilde{b}_{3k}|^2 \cos^2(\theta_{kb_3} - 2 \theta_{kg})}{1-4\left[|\tilde{b}_{3k}|^2 \cos^2(\theta_{kb_3} - 2 \theta_{kg})+ |\tilde{C}|^2 \cos^2(\theta_{kC} - \theta_{kf} -\theta_{kg}) [1+ (-1)^{1-N}\;2  \;|\tilde{b}_{3k}|^2 \cos^2(\theta_{kb_3} - 2 \theta_{kf})]\right]}.\\
\end{eqnarray*}

\subsubsection{Tensor-Tensor}
\begin{eqnarray*}
&&Case \;\;1 : \\
&&\langle h^+_{\vec{k}} h^+_{-\vec{k}} \rangle = \\
&&\frac{|g_k|^2}{\beta^2} \frac{1-4 |\tilde{C}|^2 \cos^2(\theta_{C} - \theta_{kf} - \theta_{kg})}{1-4\left[|\tilde{b}_{1k}|^2 \cos^2(\theta_{kb_1} - 2 \theta_{kg})+2 |\tilde{C}|^2 \cos^2(\theta_{kC} - \theta_{kf} -\theta_{kg}) [1+ (-1)^M \;2  \;|\tilde{b}_{1k}|^2 \cos^2(\theta_{kb_1} - 2 \theta_{kg})]\right]},\\\\
&&\langle h^{\times}_{\vec{k}} h^{\times}_{-\vec{k}} \rangle =\\
&& \frac{|g_k|^2}{\beta^2} \frac{1-4 |\tilde{C}|^2 \cos^2(\theta_{C} - \theta_{kf} - \theta_{kg})}{1-4\left[|\tilde{b}_{1k}|^2 \cos^2(\theta_{kb_1} - 2 \theta_{kg})+2 |\tilde{C}|^2 \cos^2(\theta_{kC} - \theta_{kf} -\theta_{kg}) [1+ (-1)^M \;2  \;|\tilde{b}_{1k}|^2 \cos^2(\theta_{kb_1} - 2 \theta_{kg})]\right]}.\\\\
&&Case \;\;2 : \\
&&\langle h^+_{\vec{k}} h^+_{-\vec{k}} \rangle = \\
&&\frac{|g_k|^2}{\beta^2} \frac{1-|\tilde{b}_{3k}|\cos(\theta_{kb_3}-2\theta_{kg}) -4 N |\tilde{C}|^2 \cos^2(\theta_{C} - \theta_{kf} - \theta_{kg})}{1-4\left[|\tilde{b}_{3k}|^2 \cos^2(\theta_{kb_3} - 2 \theta_{kg})+|\tilde{C}|^2 \cos^2(\theta_{kC} - \theta_{kf} -\theta_{kg}) [1+ (-1)^{1-N}\;2  \;|\tilde{b}_{3k}|^2 \cos^2(\theta_{kb_3} - 2 \theta_{kg})]\right]},\\\\
&&\langle h^+_{\vec{k}} h^+_{-\vec{k}} \rangle = \\
&&\frac{|g_k|^2}{\beta^2} \frac{1+|\tilde{b}_{3k}|\cos(\theta_{kb_3}-2\theta_{kg}) -4 (N-1)^2 |\tilde{C}|^2 \cos^2(\theta_{C} - \theta_{kf} - \theta_{kg})}{1-4\left[|\tilde{b}_{3k}|^2 \cos^2(\theta_{kb_3} - 2 \theta_{kg})+|\tilde{C}|^2 \cos^2(\theta_{kC} - \theta_{kf} -\theta_{kg}) [1+ (-1)^{1-N}\;2  \;|\tilde{b}_{3k}|^2 \cos^2(\theta_{kb_3} - 2 \theta_{kg})]\right]}.
\end{eqnarray*}

\subsubsection{Scalar-Tensor}

\begin{eqnarray*}
&&Case \;\;1 : \\
&&\langle \zeta_{\vec{k}} h^+_{-\vec{k}} + \zeta_{-\vec{k}} h^+_{\vec{k}} \rangle =\\
&& \frac{|g_k| |f_k|}{\beta \alpha} \frac{4 |\tilde{C}| \cos(\theta_{C} - \theta_{kf} - \theta_{kg})\left[2|\tilde{b}_{1k}| \cos(\theta_{kb_1}-2\theta_{kg})+ (-1)^M \right]}{1-4\left[|\tilde{b}_{1k}|^2 \cos^2(\theta_{kb_3} - 2 \theta_{kg})+ 2 |\tilde{C}|^2 \cos^2(\theta_{kb_1} - \theta_{kf} -\theta_{kg}) [1+ (-1)^M \;2  \;|\tilde{b}_{1k}|^2 \cos^2(\theta_{kb_1} - 2 \theta_{kg})]\right]},\\
&&\langle \zeta_{\vec{k}} h^{\times}_{-\vec{k}} + \zeta_{-\vec{k}} h^{\times}_{\vec{k}} \rangle =\\
&& \frac{|g_k| |f_k|}{\beta \alpha} \frac{4 |\tilde{C}| \cos(\theta_{C} - \theta_{kf} - \theta_{kg})\left[2|\tilde{b}_{1k}| \cos(\theta_{kb_1}-2\theta_{kg})+ (-1)^M \right]}{1-4\left[|\tilde{b}_{1k}|^2 \cos^2(\theta_{kb_1} - 2 \theta_{kg})+ 2 |\tilde{C}|^2 \cos^2(\theta_{kC} - \theta_{kf} -\theta_{kg}) [1+ (-1)^M \;2  \;|\tilde{b}_{1k}|^2 \cos^2(\theta_{kb_1} - 2 \theta_{kg})]\right]}.\\\\
&&Case \;\;2 : \\
&&\langle \zeta_{\vec{k}} h^+_{-\vec{k}} + \zeta_{-\vec{k}} h^+_{\vec{k}} \rangle =\\
&& \frac{|g_k| |f_k|}{\beta \alpha} \frac{4 (N-1)^2 |\tilde{C}| \cos(\theta_{C} - \theta_{kf} - \theta_{kg})\left[1-2|\tilde{b}_{3k}|\cos(\theta_{kb_3} -2\theta_{kg})\right]}{1-4\left[|\tilde{b}_{3k}|^2 \cos^2(\theta_{kb_3} - 2 \theta_{kg})+ |\tilde{C}|^2 \cos^2(\theta_{kC} - \theta_{kf} -\theta_{kg}) [1+ (-1)^{1-N}\;2  \;|\tilde{b}_{3k}|^2 \cos^2(\theta_{kb_3} - 2 \theta_{kg})]\right]},\\
&&\langle \zeta_{\vec{k}} h^{\times}_{-\vec{k}} + \zeta_{-\vec{k}} h^{\times}_{\vec{k}} \rangle =\\
&& \frac{|g_k| |f_k|}{\beta \alpha} \frac{4 N |\tilde{C}| \cos(\theta_{C} - \theta_{kf} - \theta_{kg})\left[1+2|\tilde{b}_{3k}|\cos(\theta_{kb_3} -2\theta_{kg})\right]}{1-4\left[|\tilde{b}_{3k}|^2 \cos^2(\theta_{kb_3} - 2 \theta_{kg})+ |\tilde{C}|^2 \cos^2(\theta_{kC} - \theta_{kf} -\theta_{kg}) [1+ (-1)^{1-N}\;2  \;|\tilde{b}_{3k}|^2 \cos^2(\theta_{kb_3} - 2 \theta_{kg})]\right]}.\\
\end{eqnarray*}

\subsubsection{Cross-Tensor-Tensor}

\begin{eqnarray*}
&&Case \;\;1 : \\
&&\langle h^+_{\vec{k}} h^{\times}_{-\vec{k}} \rangle =\\
&& - \frac{|g_k|^2}{\beta^2} \frac{4\left[(-1)^{1-M} 2 |\tilde{C}|^2 \cos^2(\theta_{C} - \theta_{kf} - \theta_{kg})+|\tilde{b}_{1k}| \cos(\theta_{kb_1}-2\theta_{kg})\right]}{1-4\left[|\tilde{b}_{1k}|^2 \cos^2(\theta_{kb_1} - 2 \theta_{kg})+ 2 |\tilde{C}|^2 \cos^2(\theta_{kC} - \theta_{kf} -\theta_{kg}) [1+ (-1)^M \;2  \;|\tilde{b}_{1k}|^2 \cos^2(\theta_{kb_1} - 2 \theta_{kg})]\right]},\\
&&Case \;\;2 : \\
&&\langle h^+_{\vec{k}} h^{\times}_{-\vec{k}} \rangle = 0.
\end{eqnarray*}
\endgroup
For \emph{Case 1}, $M=1$ is the case corresponding to $\tilde{C}^+ = \tilde{C}^{\times}$ (\emph{Case 1p} (plus)) while $M=0$ indicates $\tilde{C}^+ =- \tilde{C}^{\times}$ (\emph{Case 1m} (minus)). For \emph{Case 2} $N=1$ means $\tilde{C}^+ = 0$  (\emph{Case 2$\times$}) and $\tilde{C}^{\times} = \tilde{C}$ while for  $N=0$,  $\tilde{C}^{\times} = 0$ and $\tilde{C}^+ = \tilde{C}$ (\emph{Case 2$+$}).

\section{Comparison of the Different Parameter \emph{Cases}}
\begin{figure}[!htbp]
\centering
\includegraphics[scale=0.30]{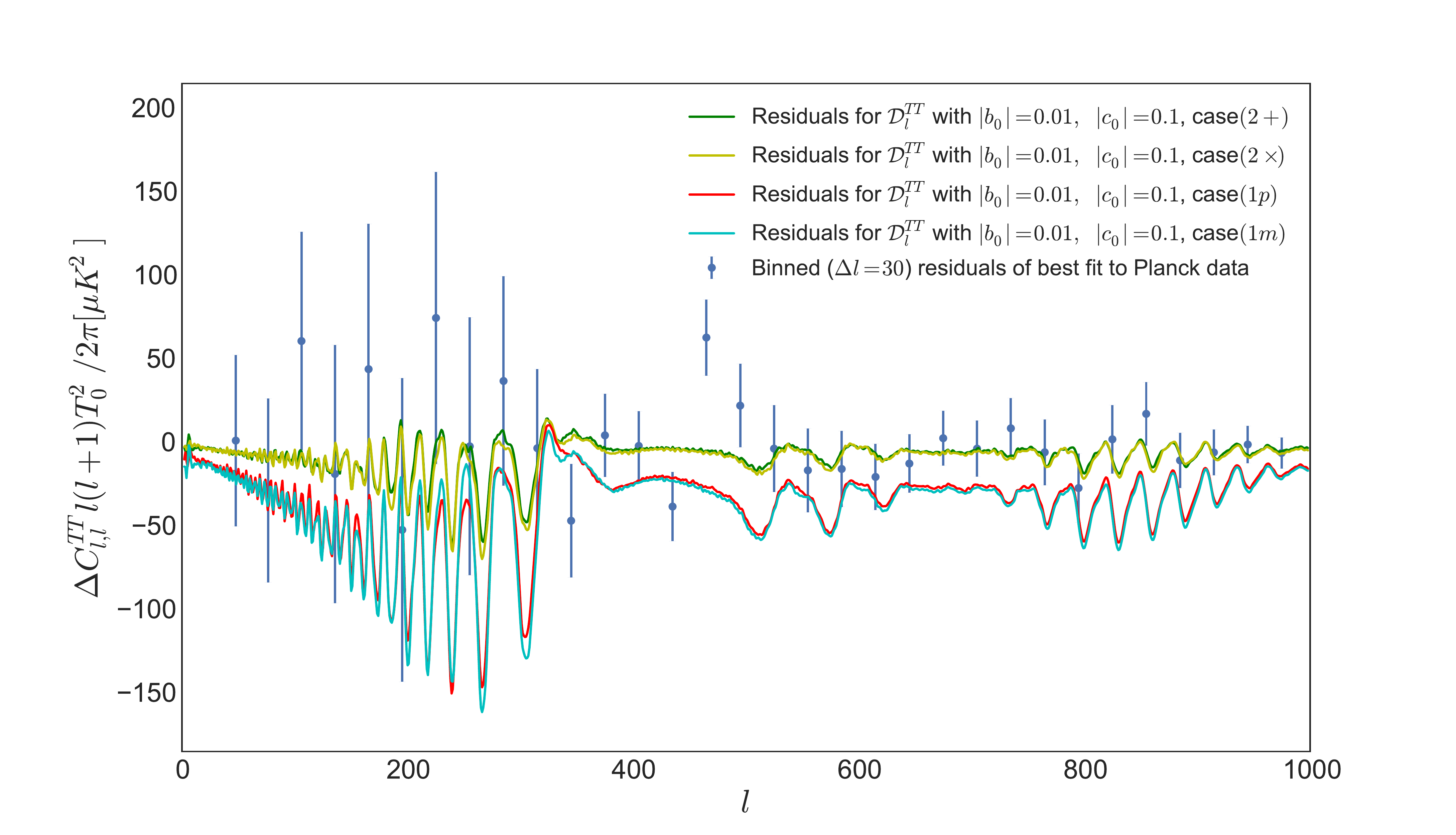}
\caption{Difference between the zero entanglement $C_l^{TT}$ and non-zero entanglement $C_l^{TT}$ for the different \emph{Cases} for $|\tilde{C}^+_{k0}|=0.1$ ($|c_0|$ on plot) and  $|\tilde{b}_{k0}| = 0.01$ ($|b_0|$ on plot), with the residual binned Planck data. The two curves with smaller oscillation amplitude are \emph{Case\;$2+$} and \emph{Case \;$2\times$}, while the larger amplitude curves correspond to \emph{Case 1p} and \emph{Case 1m}.}
\label{fig:cases1}
\includegraphics[scale=0.30]{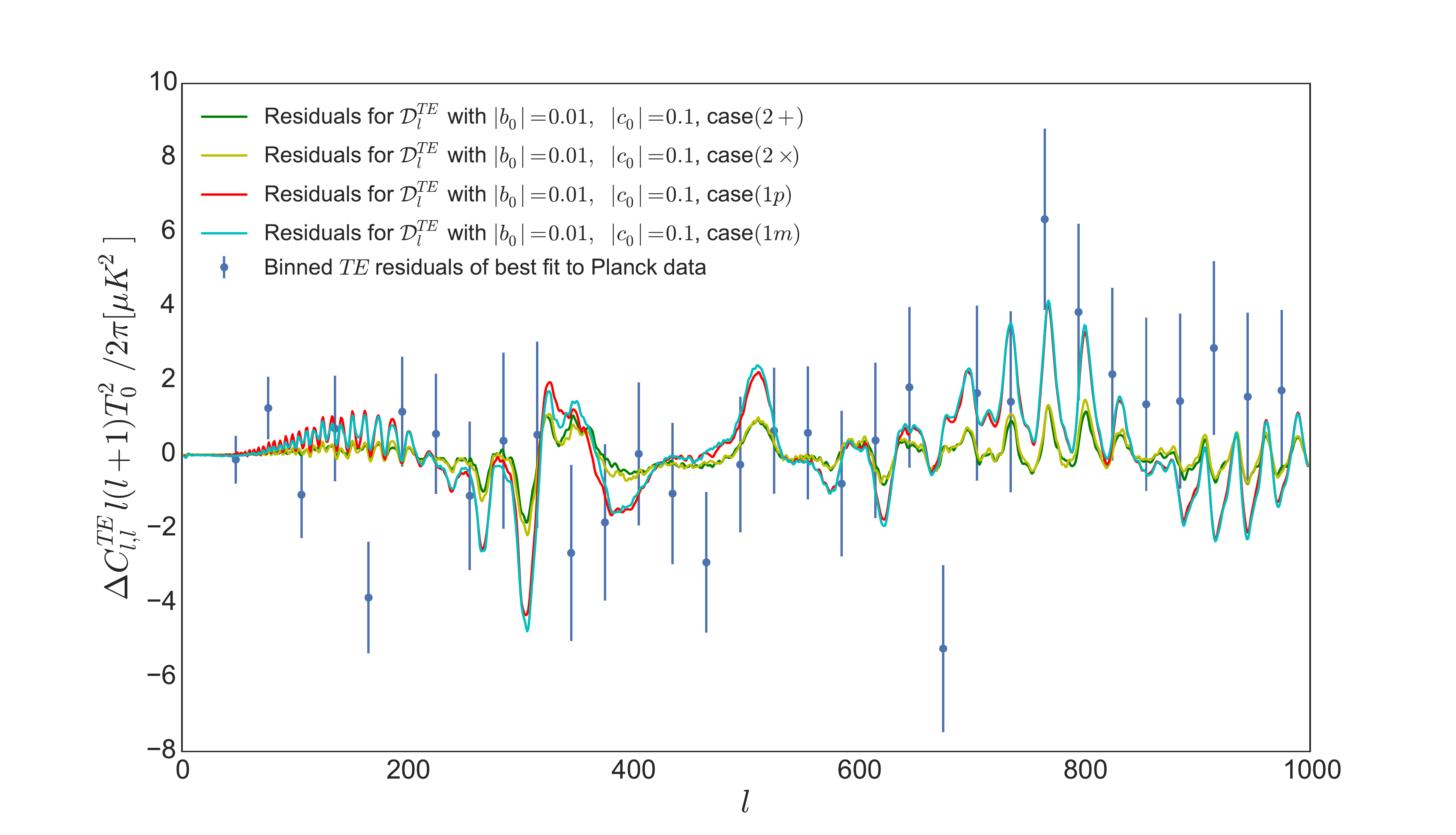}
\caption{Difference between the zero entanglement $C_l^{TE}$ and non-zero entanglement $C_l^{TE}$ for the different \emph{Cases} for $|\tilde{C}^+_{k0}|=0.1$ ($|c_0|$ on plot) and  $|\tilde{b}_{k0}| = 0.01$ ($|b_0|$ on plot), with the residual binned Planck data.The two curves with smaller oscillation amplitude are \emph{Case\;$2+$} and \emph{Case \;$2\times$}, while the larger amplitude curves correspond to  \emph{Case 1p} and \emph{Case 1m}.}
\label{fig:cases2}
\end{figure}

\begin{figure}[!htbp]
\begin{center}
\includegraphics[scale=0.30]{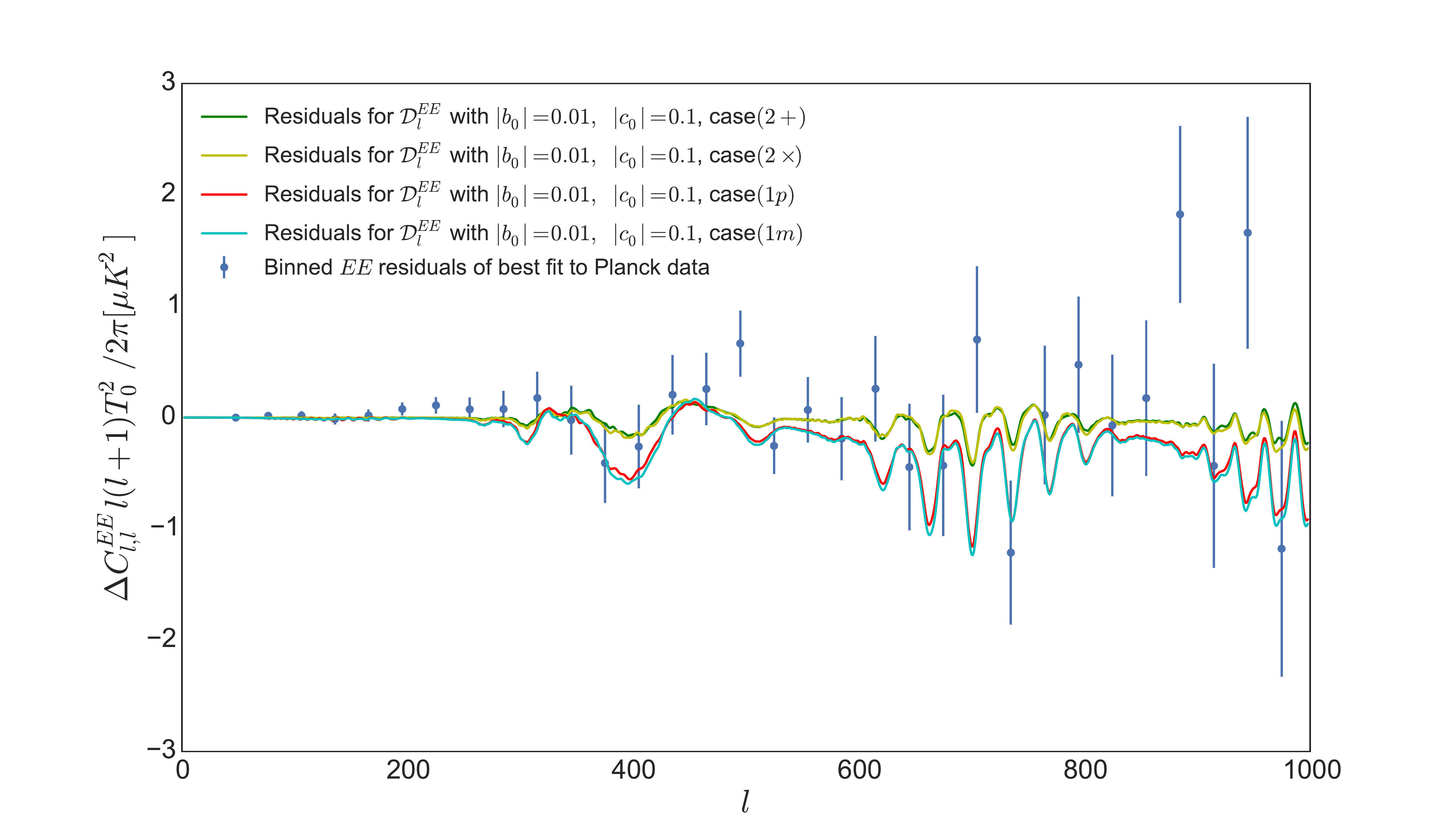}
\caption{Difference between the zero entanglement $C_l^{EE}$ and non-zero entanglement $C_l^{EE}$ for the different \emph{Cases} for $|\tilde{C}^+_{k0}|=0.1$ ($|c_0|$ on plot) and  $|\tilde{b}_{k0}| = 0.01$ ($|b_0|$ on plot), with the residual binned Planck data. The two curves with smaller oscillation amplitude are \emph{Case\;$2+$} and \emph{Case \;$2\times$}, while the larger amplitude curves correspond to  \emph{Case 1p} and \emph{Case 1m}.}
\label{fig:cases3}
\end{center}
\end{figure}

 In both temperature and polarization power spectra, for a given set of entanglement parameters,  \emph{Case \;1p} and \emph{Case \;1m} ($\tilde{C}^+ = \pm\tilde{C}^{\times}$)  exhibits larger oscillation amplitudes then  \emph{Case $2+$} and \emph{Case $2\times$}  (see figs.(\ref{fig:cases1}, \ref{fig:cases2}, \ref{fig:cases3})). 

\clearpage

\section{Von Neumann Entropy}
To calculate the von Neumann entropy $S_{\text{ent}} = - \text{Tr}(\rho_{\text{red}} \text{ln}\rho_{\text{red}} )$ it is easiest to take the trace in the eigenbasis of the reduced density matrix which can be found by solving the eigenvalue equation:

\begin{equation}
\int dx^{\prime} \rho_{\text{red}}(x, x^{\prime}) {f}_n(x^{\prime}) = p_n {f}_n(x).
\end{equation}
We find the following eigenfuncitons and eigenvalues \cite{Srednicki:1993im}:
\begin{eqnarray}
&&f_n(x) = \left(\frac{\alpha_R}{\pi}\right)^{\frac{1}{4}} H_n(\sqrt{\alpha_R} x) \exp \left(-\frac{1}{2} (\alpha_R + i \alpha_I) x^2 \right),\\
&&p_n = (1-\xi)\xi^n,
\end{eqnarray}
with $\alpha_I =\gamma_I, \alpha_R^2 = \gamma_R^2 - \beta^2, \xi = \frac{\beta}{\gamma_R +\alpha_R}$. This leads to
\begin{equation}
S_{\text{ent}} = - \sum_{n=0}^{\infty} p_n \text{ln} p_n = - \text{ln}(1-\xi) - \frac{\xi}{1-\xi} \text{ln}\xi.
\end{equation}
In terms of the mode functions $\chi$ and $\psi$ and the entanglement constant $\lambda$, $\xi$ becomes:
\begin{equation}
\xi = \frac{2 \lambda^2 }{1- 2 \lambda^2 \cos 2 (\theta_{\chi} + \theta_{\psi}) + \sqrt{1-4 \lambda^2  \cos 2 (\theta_{\chi} + \theta_{\psi}) - 4\lambda^4 \sin^2  \cos 2 (\theta_{\chi} + \theta_{\psi}) }}.
\end{equation} 
Notice that all phase information is multiplied by a power of the entanglement constant such that the amplitude of the oscillations vanish as it approaches zero. Finally looking at this limit ($\lambda \rightarrow 0$) we also see the entanglement vanishes: $$\lim_{\lambda \to\ 0} S_{\text{ent}} = 0.$$

 \section{Heisenberg Picture and Bogoliubov Transformation}
 Reference \cite{Kanno:2014} discusses the origin of oscillations using the Heisenberg picture and by describing our entangled state for two scalar fields in terms of a Bogoliubov rotation in field space. They start with two massive scalar fields in a de Sitter background with action:
 \begin{equation}
 S= \frac{1}{2} \int  d\eta \sum_{\bf{k}} a^2\left[\phi_{\bf{k}}^{\prime}\phi_{-\bf{k}}^{\prime} - (k^2 -a^2 m_{\phi}^2)  \phi_{\bf{k}}\phi_{-\bf{k}} +\chi_{\bf{k}}^{\prime}\chi_{-\bf{k}}^{\prime} - (k^2 -a^2 m_{\chi}^2)\chi_{\bf{k}}\chi_{-\bf{k}})\right],
 \end{equation}
 which can be expanded with raising and lowering operators:
 \begin{eqnarray}
&\phi_{\bf{k}} = a_{\bf{k}}\; u_k(\eta) + a_{-\bf{k}}^{\dagger} \;u^*_k(\eta), \quad \quad& [a_{\bf{k}}, a^{\dagger}_{\bf{p}}]=\delta_{\bf{k}\bf{p}},\\
& \chi_{\bf{k}} = b_{\bf{k}}\; v_k(\eta) + b_{-\bf{k}}^{\dagger} \;v^*_k(\eta), \quad \quad& [b_{\bf{k}}, b^{\dagger}_{\bf{p}}]=\delta_{\bf{k}\bf{p}}.\\ 
  \end{eqnarray}
  The $u_k(\eta)$ and $v_k(\eta)$ are the Bunch-Davies vacuum mode functions for $\phi_{\bf{k}}$ and $\chi_{\bf{k}}$ and the annihilation operators $a_{\bf{k}}$ and $b_{\bf{k}}$ annihilate their respective BD vacuums. 
  \begin{equation}
  a_{\bf{k}} |0\rangle_{\phi}^{BD} = 0, \quad \quad  b_{\bf{k}} |0\rangle_{\chi}^{BD} = 0  .
  \end{equation} 
  Next they suppose that an entangled Gaussian state of the same form as our entangled state between two scalars (Eqn 2.6 in \cite{Albrecht:2014eaa}) is annihilated by new annihilation operators, 
 \begin{equation}
  \tilde{a}_{\bf{k}} |\Psi \rangle =  \tilde{b}_{\bf{k}} |\Psi \rangle = 0  ,
  \end{equation} 
 defined by a Bogoliubov transformation that mix the BD raising and lowering operators:
 \begin{equation}
 \tilde{a}_{\bf{k}} = \alpha_k a_{\bf{k}} + \beta_k b^{\dagger}_{\bf{k}},  \quad\quad  \tilde{b}_{\bf{k}} = \alpha_k b_{\bf{k}} + \beta_k a^{\dagger}_{\bf{k}}.
 \end{equation}
 The Bogoliubov coefficients obey the usual expression:
 \begin{equation}
 |\alpha_k|^2-|\beta_k|^2 = 1.
 \end{equation}
 The two-point function for one of the fields, in terms of the Bogoliubov coefficients takes the general form:
 \begin{equation}
 \langle \Psi| \phi_k \phi_{-k}|\Psi \rangle =  |u_k|^2 (1+ 2 |\beta_k|^2).
 \end{equation}
 
 By requiring that $\tilde{a}_{\bf{k}}$ and  $\tilde{b}_{\bf{k}}$ annihilate a Gaussian state of our form (Eqn 2.6 in \cite{Albrecht:2014eaa}), they get expressions for the coefficients $A_k,  B_k$ and $C_k$ in terms of the BD mode functions and Bogoliubov coefficients. In their calculation the time dependence of $A_k,  B_k$ and $C_k$ depends solely on the BD mode functions produced by varying the decoupled action of the fields. The annihilation operators $\tilde{a}_{\bf{k}}$ and  $\tilde{b}_{\bf{k}}$ are not time dependent. Annihilating the Gaussian entangled state with such annihilation operators does not impose any time evolution on the state (or the coefficients  $A_k,  B_k$ and $C_k$), and in particular does not give the time evolution we get by applying the Schr\"odinger equation to our state. Since the time dependence of the Bogoliubov transformed state is solely determined by varying the decoupled action above, the result is the same time dependence as regular BD modes. In this paper and our previous paper \cite{Albrecht:2014eaa}, we work within the framework of finite inflation and explicitly do not start with BD vacuum. Our state is therefore not  equivalent to the Bogoliubov transformed state in \cite{Kanno:2014}. 
 
 To illustrate this difference better we present a simple toy model\footnote{ We thank Jiro Soda for useful communications on this topic including his proposition of this toy model.}. We are interested in finding the two point function $\langle \Psi| \phi_k \phi_{-k}|\Psi \rangle$ of a field in a general state $|\Psi\rangle$ which gets annihilated by Bogoliubov rotated annihilation operators described in Eqn A.11. For the purposes of our toy model we consider fields whose mode functions are linear superpositions of BD mode functions, in terms of $k$-dependent coefficients $\tilde{A}_k$ and $\tilde{B}_k$. This is to model the fact that our state is an excited state:
 
 \begin{equation}
 u_k = \tilde{A}_k u^{BD}_k + \tilde{B}_k u^{BD*}_k.
 \end{equation}
 Plugging this definition of $u_k$ into the expression for the two point function (Eqn A.14) we find:
 \begin{equation}
\langle \Psi | \phi_k \phi_{-k} |\Psi \rangle =  \left( |\tilde{A}_k|^2 +|\tilde{B}_k|^2 + 2 |\tilde{A}_k| |\tilde{B}_k| \cos(\theta^A_k +\theta^B_k + 2 \theta^{BD}_k)\right) |u_k^{BD}|^2 \left( 1 + 2 |\beta_k|^2\right),
  \end{equation}
   where $\theta^A_k, \theta^B_k, \theta^{BD}_k$ are the $k$-dependent phases for $\tilde{A}_k, \tilde{B}_k, u^{BD}_k$ respectively. By setting $\tilde{B}_k = 0$ and $\tilde{A}_k=1$ we recover the original BD Bogoliubov rotated solution from  \cite{Kanno:2014}. However, when the field is in the excited state the two-point function will have k-dependent oscillatory behavior induced by the cosine of the $k$-dependent phases. This toy model demonstrates that while the Bogoliubov transformation from \cite{Kanno:2014} alone does introduce mixing between the field modes it does not provide the same mixing and oscillatory behavior as can be found in more general states (as given by Eqn A.15 in the toy model).  We see the same phenomenon in the comparison of our full results with the results in \cite{Kanno:2014}.
 
 \clearpage

\bibliographystyle{JHEP}
\bibliography{TensorScalarEntanglement}
\end{document}